\documentclass[prb,amsmath,amssymb,amsbsy,twocolumn,superscriptaddress,showpacs]{revtex4-1}

\usepackage[svgnames]{xcolor}
\definecolor{darkblue}{RGB}{0 60 120}
\definecolor{eggplant}{RGB}{190 10 150}
\definecolor{darkgray}{RGB}{70 70 70}
\usepackage{amsfonts}
\usepackage{amsmath}
\usepackage{amssymb}
\usepackage{bm}
\usepackage{graphicx}
\usepackage[breaklinks, colorlinks = true,linkcolor=eggplant,urlcolor=darkblue,citecolor=eggplant]{hyperref}
\usepackage{mathrsfs}
\usepackage[lofdepth,lotdepth,caption=false]{subfig}
\usepackage{varwidth}
\usepackage{wrapfig}
\usepackage{times}
\usepackage{bm}
\usepackage[abs]{overpic}
\usepackage{longtable}
\usepackage{multirow}
\usepackage{tikz}
\newcommand{\liiro}{Li${}_{2}$IrO${}_{3}$}

\newcommand{\ir}{$\text{Ir}^{4+}$}
\newcommand{\jhalf}{$j_{\text{eff}}=1/2$}
\newcommand{\jhalftitle}{$\boldmath{j_{\text{eff}}=1/2}$}
\newcommand{\jthreehalf}{$j_{\text{eff}}=3/2$}
\newcommand{\hhc}{hyperhoneycomb}
\newcommand{\harzero}{$\mathcal{H}\text{--}0$}
\newcommand{\har}{$\mathcal{H}\text{--}1$}
\newcommand{\ospa}[1]{SP$_{a^{#1}}$}
\newcommand{\ospb}[1]{SP$_{b^{#1}}$}
\newcommand{\ospabar}[1]{$\overline{\text{SP}}_{a^{#1}}$}
\newcommand{\ospbbar}[1]{$\overline{\text{SP}}_{b^{#1}}$}
\newcommand{\rdot}{\begin{tikzpicture}\filldraw[color=red!50, fill=red!0, very thick](0,0) circle (.15);\end{tikzpicture}}
\newcommand{\bdot}{\begin{tikzpicture}\filldraw[color=blue!50, fill=red!0, very thick](0,0) circle (.15);\end{tikzpicture}}
\newcommand{\gdot}{\begin{tikzpicture}\filldraw[color=green!50, fill=red!0, very thick](0,0) circle (.15);\end{tikzpicture}}

\begin{document}

\title{Theory of Magnetic Phase Diagrams in Hyperhoneycomb and
  Harmonic-honeycomb Iridates}

\author{Eric Kin-Ho Lee}
\affiliation{Department of Physics and Center for Quantum Materials,
University of Toronto, Toronto, Ontario M5S 1A7, Canada.}
\author{Yong Baek Kim}
\affiliation{Department of Physics and Center for Quantum Materials,
University of Toronto, Toronto, Ontario M5S 1A7, Canada.}
\affiliation{Canadian Institute for Advanced Research/Quantum Materials Program, Toronto, Ontario MSG 1Z8, Canada}
\affiliation{School of Physics, Korea Institute for Advanced Study, Seoul 130-722, Korea.}

\begin{abstract}
  Motivated by recent experiments, we consider a generic spin model in
  the \jhalf{} basis for the hyperhoneycomb and harmonic-honeycomb
  iridates. Based on microscopic considerations, the effect of an
  additional bond-dependent anisotropic spin exchange interaction
  ($\Gamma$) beyond the Heisenberg-Kitaev model is investigated.  We
  obtain the magnetic phase diagrams of the hyperhoneycomb and
  harmonic-honeycomb (\har{}) lattices via a combination of the
  Luttinger-Tisza approximation, single-$\mathbf{Q}$ variational
  ansatz, and classical Monte Carlo simulated annealing. The resulting
  phase diagrams on both systems show the existence of incommensurate,
  non-coplanar spiral magnetic orders as well as other commensurate
  magnetic orders. The spiral orders show counter-propagating spiral
  patterns, which may be favorably compared to recent experimental
  results on both iridates. The parameter regime of various magnetic
  orders and ordering wavevectors are quite similar in both systems.
  We discuss the implications of our work to recent experiments and
  also compare our results to those of the two dimensional honeycomb
  iridate systems.
\end{abstract}
\date{\today}
\maketitle

\section{\label{sec:intro}Introduction}
Iridium-based compounds\cite{yanagishima2001metal,
  nakatsuji2006metallic, matsuhira2007metal,
  singh2010antiferromagnetic, liu2011long, singh2012relevance,
  choi2012spin, qi2012strong} have spurred considerable interest
because the interplay of strong atomic spin-orbit coupling (SOC),
electronic correlations, crystal field effects, and sizable orbital
overlaps in these compounds have been shown to give rise to exotic
states of matter ranging from topological insulators to Weyl
semimetals and quantum spin liquids.\cite{hasan2010topological,
  qi2011topological, hasan2011three, pesin2010mott,
  yang2010topological, you2012doping, witczak2013correlated} Drawing
significant attention is the A$_2$IrO$_3$ family of layered honeycomb
iridates\cite{jackeli2009mott, shitade2009quantum,
  chaloupka2010kitaev, singh2010antiferromagnetic, liu2011long,
  kimchi2011kitaev, bhattacharjee2012spin, singh2012relevance,
  ye2012direct, lovesey2012magnetic, choi2012spin, comin2012na,
  kim2012topological, mazin2012na, schaffer2012quantum,
  foyevtsova2013ab, clancy2012spin, gretarsson2013crystal,
  gretarsson2013magnetic, chaloupka2013zigzag, cao2013evolution}:
these materials were first described by the Kitaev
model\cite{jackeli2009mott} and later by the Heisenberg-Kitaev (HK)
model\cite{chaloupka2010kitaev}, both of which hosts a $Z_2$ quantum
spin-liquid ground state first discovered in the context of Kitaev's
exactly solvable spin-1/2 model.\cite{kitaev2006anyons} Subsequently,
similar exactly solvable spin models on several other two-dimensional
(2D) and three-dimensional (3D) lattices were
investigated.\cite{yang2007mosaic, yao2007exact, si2008anyonic,
  mandal2009exactly, chua2011exact, hermanns2014quantum}

Experimentally, the sodium honeycomb iridate---Na$_2$IrO$_3$---has
been shown to order magnetically in the zigzag
phase\cite{liu2011long,ye2012direct,choi2012spin}, while the lithium
counterpart---\liiro{}---is believed to also order magnetically but in
a different ground state.\cite{cao2013evolution} The quest to find an
appropriate minimal model that can explain the observed orders, low
energy excitations, as well as bulk thermodynamic properties has been
actively pursued.  To address these points, localized pseudospin
models involving exchanges in addition to the Heisenberg and Kitaev
couplings have been considered, including the symmetry-allowed
off-diagonal exchange\cite{rau2014generic, katukuri2014kitaev,
  yamaji2014honeycomb}, further neighbour
exchanges\cite{kimchi2011kitaev, reuther2014spiral}, and
distortion-induced anisotropies\cite{katukuri2014kitaev,
  yamaji2014honeycomb}.  Electronic band structure
calculations\cite{kim2012topological, mazin2012na, foyevtsova2013ab,
  kim2013strain} have also been investigated to further our
understanding of these compounds.  Indeed, a suitable minimal model
may predict the behaviour of these layered honeycomb iridates under a
variety of perturbations and may direct us to the discovery of exotic
physics, including the illusive quantum spin liquid phase.

Recently, two three-dimensional polymorphs of the layered honeycombs
with chemical formula \liiro{} have been discovered, fuelling intrigue
over the family of honeycomb-based iridate compounds. Curiously, these
two materials---the \hhc{}
($\mathcal{H}\text{--}0$)\cite{takayama2014hyper} and
harmonic-honeycomb (\har{})\cite{modic2014realization}--- behave
rather similarly: they have the same stoichiometry as the layered
honeycomb with analogous three-dimensional crystal structures, they
order magnetically with no net moment at similar temperatures
($\sim\negmedspace38~\text{K}$), and their magnetization curves behave
non-linearly at low fields.  Furthermore, thermodynamic
measurements\cite{takayama2014hyper,modic2014realization} suggested
that both 3D iridates might order unconventionally in a non-collinear
fashion and recent single-crystal resonant magnetic x-ray diffraction
experiments\cite{biffin2014non,biffin2014unconventional} find a
complex incommensurate structure with non-coplanar and
counter-propagating moments in both iridates. Indeed, with the
discovery of these two 3D honeycomb compounds and to build on past
work in both the intermediate-coupling regime\cite{lee2014topological}
and localized\cite{mandal2009exactly, lee2014heisenberg, lee2014order,
  nasu2014finite, lee2014emergent, kimchi2013three,
  nasu2014vaporization} limits, the search for a relevant minimal
model to describe the family of honeycomb iridates has now become more
fascinating.  Can the same minimal model describe both the 2D and 3D
honeycomb iridates while capturing the similarities and heterogeneity
amongst these compounds?  In this work, driven by the fact that these
compounds have similar local structures, we suggest that the
nearest-neighbor $J\text{-}K\text{-}\Gamma$ model, which was recently
used to describe the layered honeycomb iridates\cite{rau2014generic},
is a good starting point in the discussion of this family of
compounds, especially in regards to the magnetic order.

The rest of the paper is organized as follows. We begin, in Sec.
\ref{sec:lattices}, by discussing the \hhc{} and \har{} lattices,
drawing special attention to the similarities and differences between
these three-dimensional analogues of the honeycomb lattice. In Sec.
\ref{sec:model}, we describe the minimal effective model of \jhalf{}
pseudospins that describes the low-energy physics of the \hhc{} and
\har{} lattices in the Mott insulating limit. The presence of the
off-diagonal exchange term, $\Gamma$, along with the familiar
Heisenberg and Kitaev terms, is shown to be necessary from a
strong-coupling expansion of the underlying, multi-band electronic
model. Following this, in Sec. \ref{sec:pd} we present the classical
phase diagram of these models using a combination of Luttinger-Tisza
approximation, single-$\mathbf{Q}$ ansatz minimization, and simulated
annealing.  We show that the \hhc{} and \har{} lattices have similar
phase diagrams, but differ from the phase diagram of the 2D honeycomb
iridates in important ways.  In addition to analogues of the
ferromagnetic, antiferromagnetic, zigzag, and stripy phases of the HK
model, we find a number of non-coplanar, counter-propagating spiral
phases, multiple-$\mathbf{Q}$ states, as well as phases that have no
direct analogues in the HK-limit.  We expound these phases and their
static structure factors in Sec. \ref{sec:orders}, thus providing
results that can be compared with experiments. Finally, in
Sec. \ref{sec:discussion}, we discuss the relevance of our results to
the newly discovered 3D-\liiro{} compounds.

\section{\label{sec:lattices}Structure of \hhc{} and \har{} lattices}
\begin{figure}[h!]
  \centering
  \setlength\fboxsep{0pt}
  \setlength\fboxrule{0pt}
  \subfloat[][Hyperhoneycomb lattice]{
    \label{fig:hhc_lattice}
    \fbox{\begin{overpic}[scale=1,clip=true,trim=15 0 -15 0]{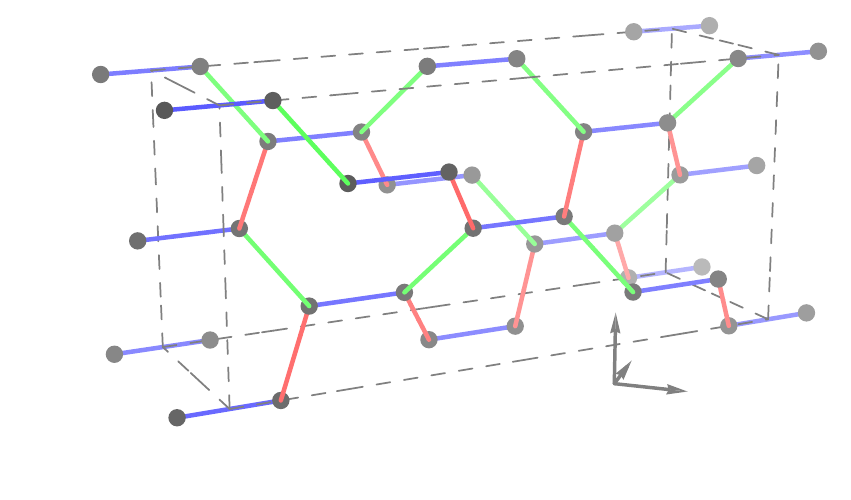}
        \put(36,30){$a$}
        \put(45,60){$b$}
        \put(133,32){$c$}
        \put(185,29){$\hat{x}$}
        \put(167,41){$\hat{y}$}
        \put(160,54){$\hat{z}$}
        \put(72,40){$x$}
        \put(58,60){$y$}
        \put(48,19){$z$}
      \end{overpic}}
  }

  \subfloat[][\har{} lattice]{
    \label{fig:har_lattice}
    \fbox{\begin{overpic}[scale=1,clip=true,trim=15 0 -15 0]{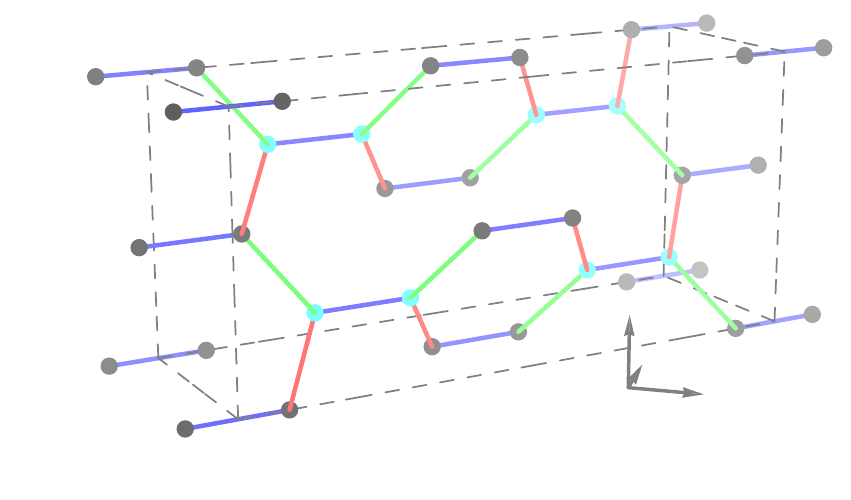}
        \put(36,28){$a$}
        \put(46,59){$b$}
        \put(133,32){$c$}
        \put(188,29){$\hat{x}$}
        \put(170,41){$\hat{y}$}
        \put(164,54){$\hat{z}$}
        \put(74,38){$x$}
        \put(59,59){$y$}
        \put(51,17){$z$}
      \end{overpic}}
  }
  \caption{\label{fig:lattices}(Color online) Ideal iridium ion
    network in the \hhc{} and \har{} lattices.  The $x$, $y$, and $z$
    bonds are colored red, green, and blue, respectively.  The
    bridging-sites are colored cyan in Fig. \ref{fig:har_lattice},
    and the $z$-bonds connecting adjacent bridging-sites are the
    bridging-$z$-bonds.  The gray dashed boxes are the conventional
    unit cells.  The conventional lattice vectors---$a$, $b$, and
    $c$---are marked accordingly.}
\end{figure}
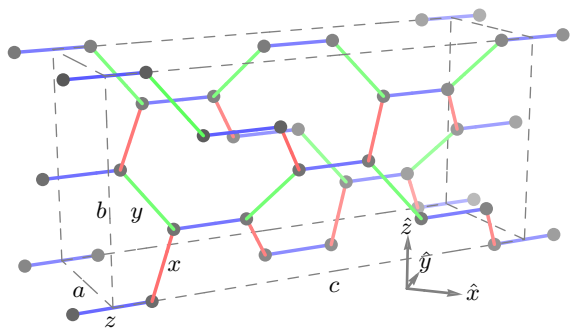
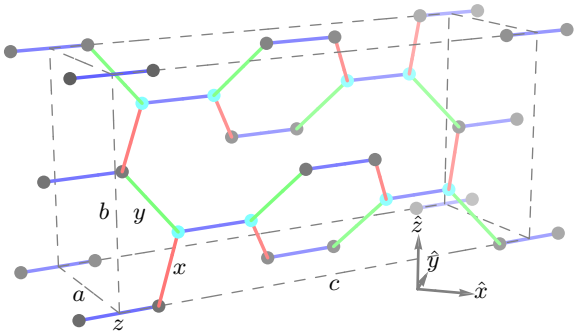

The ideal iridium ion (Ir) network in the \hhc{}
(Fig. \ref{fig:hhc_lattice}) and \har{} (Fig. \ref{fig:har_lattice})
lattices share many common features but also have some important
differences.  In this section, we highlight these main points and
introduce conventions relevant to the description of the magnetic
phases explored in Sec. \ref{sec:orders}.

The conventional unit cells of both \hhc{} and \har{} lattices are
orthorhombic with the former being
face-centered\cite{takayama2014hyper} and the latter being
base-centered\cite{modic2014realization}.  The \hhc{} primitive unit
cell has four Ir while the \har{} primitive unit cell has eight;
hence, the conventional unit cell of \hhc{} contains four primitive
units while in the \har{}, there are two.  In our chosen coordinate
system, the conventional lattice vectors of both \hhc{} and \har{}
lattices are given by the orthorhombic set $\mathbf{a}=(-2,2,0)$,
$\mathbf{b}=(0,0,4)$, and $\mathbf{c}=(6,6,0)$.  All the Ir sites
reside in their respective oxygen octahedra and these octahedra share
edges between NN Ir sites.

The Ir sites are three-fold coordinated, with the three NN bonds
labelled as $x$, $y$, and $z$ as illustrated in
Figs. \ref{fig:hhc_lattice} and \ref{fig:har_lattice}.  In the
coordinate system that we have chosen, the label represents the
direction to which the bond is normal.  For example, the sites joined
by an $x$-bond form either the vector $(0,1,1)$ or $(0,1,-1)$.  We
will see that this labelling scheme is also used in describing the
low-energy Hamiltonian of these lattices in Sec. \ref{sec:model}.

The \hhc{} (also known as the \harzero{}) and \har{} lattices have
been described as strips of layered honeycombs joined in an ABAB
fashion along the $\hat{c}$ direction.  The A and B strips are
mirrored copies of each other and are hence related by glide
planes.\cite{modic2014realization,kimchi2013three} In the
$\mathcal{H}\text{-}n$ notation, $n$ refers to the number of rows of
hexagons in the each of the honeycomb strips.  For the \hhc{} lattice,
$n=0$ indicates that no complete hexagons are present.  For the \har{}
lattice, each honeycomb strip is one row in width.  Despite the
difference in the topology of the Ir network, the projection along the
conventional planes ($ab$, $ac$, and $bc$ planes) are identical
between the two lattices.

We define a \textit{honeycomb-plane} associated with every site: it is
the plane that is spanned by the site's three NN bonds.
Alternatively, it can be defined as the plane that contains the
honeycomb strip to which the site belongs.  Since the honeycomb strips
are related by glide planes, there are only two orientations that
honeycomb-planes can take.  The honeycomb-planes are important when
discussing the magnetic orders in Sec. \ref{sec:orders}, especially for
the spiral phases.

All Ir are symmetry-equivalent in the \hhc{} lattice, but in the
\har{} lattice, there are two inequivalent sets.  These two
inequivalent sets of Ir in the \har{} lattice can be classified by
considering whether the site shares the same honeycomb-plane with all
its NNs.  We denote the sites that don't share the same
honeycomb-plane with all its NNs as \textit{bridging-sites} since they
connect the A and B honeycomb strips.  Consequently, we call the
$z$-bonds that connect bridging-sites \textit{bridging-$z$-bonds},
while the other symmetry-inequivalent $z$-bonds
\textit{non-bridging-$z$-bonds}.  This terminology is used in the
discussion of the spin patterns in several of the magnetic orders.

\section{\label{sec:model}\jhalftitle{} generic pseudospin model}
In compounds containing \ir{} ions residing in oxygen octahedral
crystal fields, the combination of large atomic spin-orbit coupling
and strong electronic correlations can result in low-energy physics
describable by highly anisotropic \jhalf{} pseudospin models
(hereafter, we will use \textit{spin} and \textit{pseudospin}
interchangeably for brevity).\cite{kim2008novel, jackeli2009mott,
  nussinov2013compass, rau2014generic} When the bonds are comprised of
edge-shared octahedra, such as the NN bonds found in both the \hhc{}
and the \har{} lattices, a symmetric off-diagonal exchange
term---denoted by $\Gamma$---is generically present in addition to the
Heisenberg and Kitaev exchange terms.\cite{rau2014generic,
  katukuri2014kitaev, yamaji2014honeycomb} As such, a minimal
Hamiltonian that describes the low-energy physics of these systems can
be written as
\begin{equation}
  \label{eq:ham}
  H=\sum_{\langle i,j\rangle \in \alpha \beta (\gamma)}
  \left[J \mathbf{S}_i \cdot \mathbf{S}_j
    + K S^{\gamma}_{i} S^{\gamma}_{j}
    \pm \Gamma \left(S^{\alpha}_{i} S^{\beta}_{j} + S^{\beta}_{i} S^{\alpha}_{j}\right)\right],
\end{equation}
where $\mathbf{S}_i$ is the \jhalf{} pseudospin at site $i$, the
summation is over NN bonds $\langle i, j \rangle$ labelled by
$\gamma\in(x,y,z)$, and $\langle i, j \rangle \in \alpha \beta
(\gamma)$ is shorthand for $\langle i, j \rangle \in \gamma, \alpha
\neq \beta \neq \gamma$.  The $\pm$ sign in front of $\Gamma$ is a
reminder that, unlike the $J$ and $K$ terms, the $\Gamma$ term can
have relative minus signs on different bonds.  The sign structures of
$\Gamma$ for both the \hhc{} and \har{} lattices are shown in
Fig. \ref{fig:sign_lattice} (see Appendix \ref{app:sign} for details
on the choice of the $\Gamma$ sign structure).

The exchange interactions $J$, $K$, and $\Gamma$ can be obtained
microscopically via a strong coupling expansion of a $t_{2g}$
multi-band tight-binding model with on-site interactions and large
atomic SOC.  The details can be found in Appendix \ref{app:model} and
Ref. \onlinecite{rau2014generic}.  Here we summarize the exchange
interactions' dependence on the parameters of the multi-band Hubbard
model, which include Coulomb repulsion $U$, Hund's coupling $J_H$,
spin-orbit coupling $\lambda$, and three $t_{2g}$ hopping amplitudes
denoted by $t_1$ through $t_3$
\begin{align}
  \label{eq:exchanges}
  J&= \frac{4}{27}\left[
    \frac{(2t_1+t_3)^2(4J_{H}+3U)}{U^2} - \frac{16J_{H}(t_1-t_3)^2}{(2U+3\lambda)^2}\right]\nonumber \\
  K&= \frac{32J_H}{9}\left[
    \frac{(t_1-t_3)^2-3t^2_2}{(2U+3\lambda)^2}\right],\  \Gamma= \frac{64J_H}{9}
    \frac{t_2(t_1-t_3)}{(2U+3\lambda)^2}.
\end{align}
We note that all three exchange interactions are present at this order
of perturbative expansion and that the presence of Hund's coupling
$J_H$ is needed for a finite $K$ and $\Gamma$.  Furthermore, the signs
of the exchange interactions sensitively depend on relative magnitudes
and signs of $t_1$ through $t_3$, which themselves are functions of
Slater-Koster amplitudes\cite{slater1954simplified} that parametrize
$t_{2g}$ orbital overlaps and oxygen-iridium orbital overlaps (details
can be found in Appendix \ref{app:model} and Ref.
\onlinecite{rau2014generic}).  Generically, this sensitivity implies
that all the phases we find are potentially experimentally relevant to
the 3D-\liiro{} compounds.

\section{\label{sec:pd}Construction of phase diagrams}
The NN \jhalf{} pseudospin Hamiltonian in Eq. \ref{eq:ham} describes
the local, low-energy physics of the \hhc{} and \har{} lattices in the
strong SOC and electron correlation limit.  Unlike the 2D honeycomb
iridates, both the \hhc{} and the \har{} lattices are
three-dimensional, thus making numerical techniques for quantum spin
models, such as exact diagonalization and density matrix
renormalization group, challenging to implement.  This is further
exacerbated by the large unit cells in these 3D lattices: compared to
the two Ir ions in the 2D-honeycomb lattice's primitive unit cell,
there are four and eight Ir ions in the primitive unit cell for the
\hhc{} and \har{} lattices, respectively.

Although these numerical issues in the quantum analysis are difficult
to overcome, a classical understanding of these three-dimensional
pseudospin models is both important and relevant insomuch that the
\hhc{} and \har{} iridates are believed to be
magnetically-ordered.\cite{takayama2014hyper,modic2014realization} A
classical analysis may be sufficient since, in the 2D-honeycomb
pseudospin model, the classical phase diagram well-approximated the
magnetic phases found via exact
diagonalization.\cite{chaloupka2010kitaev, chaloupka2013zigzag,
  okamoto2013global, rau2014generic} Furthermore, quantum
fluctuations---the key ingredient in generating exotic, non-classical
phases---are generally believed to be suppressed in 3D systems
compared to 2D systems.  To this end, we analyze the \jhalf{}
pseudospin Hamiltonians of both \hhc{} and \har{} lattices in the
classical limit using a combination of the Luttinger-Tisza
approximation\cite{luttinger1946theory, litvin1974luttinger},
single-$\mathbf{Q}$ ansatz minimization, and classical Monte Carlo
simulated annealing.

\subsection{\label{subsec:methods}Methods}
In the Luttinger-Tisza approximation (LTA),\cite{luttinger1946theory,
  litvin1974luttinger} the local constraint of $|S_i|=1$ is relaxed to
a global constraint $\sum_i|S_i|=N$.  The lowest energy configuration
satisfying the global constraint can be easily solved in momentum
space.  This configuration is then checked against the local
constraint: if it is satisfied, then the LTA succeeded in finding the
exact classical ground state; otherwise, the LTA failed but has
instead found a lower bound of the ground state energy.  In the
parameter regions where LTA failed, we implemented a combination of
single-$\mathbf{Q}$ ansatz minimization and classical Monte Carlo
simulated annealing to find the classical ground state configuration.

In the single-$\mathbf{Q}$ ansatz minimization, an upper bound to the classical
energy is determined variationally by a trial spin configuration of
the form
\begin{equation}
  \label{eq:singleQAnsatz}
  \mathbf{S}_a(\mathbf{r}_i)=\mathbf{\hat{e}}^{z}_{a}\cos \alpha_a  +
  \sin \alpha_a \left[
    \mathbf{\hat{e}}^{x}_a\cos(\mathbf{Q}\cdot\mathbf{r}_i) +
    \mathbf{\hat{e}}^{y}_a\sin(\mathbf{Q}\cdot\mathbf{r}_i)
  \right],
\end{equation}
where $\mathbf{S}_a(\mathbf{r}_i)$ is the pseudospin on sublattice $a$
in the (conventional) unit cell $\mathbf{r}_i$,
$\mathbf{\hat{e}}^{i}_{a}$ is an orthonormal set of vectors defining a
local frame for sublattice $a$, $\alpha_a$ is the canting angle for
sublattice $a$, and $\mathbf{Q}$ is the $\mathbf{Q}$-vector common to
all sublattices. The energy of this ansatz is minimized to find a
variational bound to the classical ground state energy.  Since the
wavevectors $\mathbf{Q}$ and
$\mathbf{Q}'\equiv\mathbf{g}\pm\mathbf{Q}$ (where $\mathbf{g}$ is a
reciprocal lattice vector) parametrize the same set of spin
configurations, we can restrict the domain over which minimization of
$\mathbf{Q}$ is performed to the first Brillouin zone.

Classical simulated annealing using the single-spin Metropolis
algorithm and periodic boundary conditions was performed in
conjunction with the single-$\mathbf{Q}$ ansatz minimization.
Simulated annealing has the advantage over single-$\mathbf{Q}$ ansatz
minimization in that it can access states that are characterized by
multiple $\mathbf{Q}$-vectors and is less prone to being trapped by
local minima in energy.  On the other hand, although simulated
annealing can access single-$\mathbf{Q}$ states given by
Eq. \ref{eq:singleQAnsatz}, the allowable $\mathbf{Q}$-vectors must be
commensurate with the finite system sizes chosen in our simulations.
A saw-tooth like temperature annealing profile was used to better
traverse the spin-configuration space and a minimum of $1\times10^{6}$
sweeps (updates per spin) were used for each simulation.

Over 8000 $(J,K,\Gamma)$ parameter points for each lattice were
analyzed, with particular focus on spiral phases.  For each parameter
point, simulated annealing was performed with random initial
conditions on systems with $n\times n\times n$ conventional unit cells
where $n\le10$.  For the spiral regions, we performed additional
annealing on systems with $m\times 1 \times 1$ conventional unit
cells, where $m\le100$ is the direction along the spiral wavevector.
In addition, a minimum of $1000$ single-$\mathbf{Q}$ ansatz
minimization runs were also performed per parameter point.  The
minimum energy amongst these simulated annealing and
single-$\mathbf{Q}$ ansatz minimization runs is deemed the variational
bound of the ground state energy and the corresponding pseudospin
configuration is used to characterize the magnetic order of the ground
state at that parameter point.

\subsection{\label{subsec:generalconsiderations}General considerations}
\begin{figure}[htbp!]
  \centering
  \setlength\fboxsep{0pt}
  \setlength\fboxrule{0pt}
  \subfloat[][Hyperhoneycomb model]{
    \label{fig:pd_hhc_pos}
    \fbox{\begin{overpic}[scale=1,clip=true,trim=0 -10 0 0]{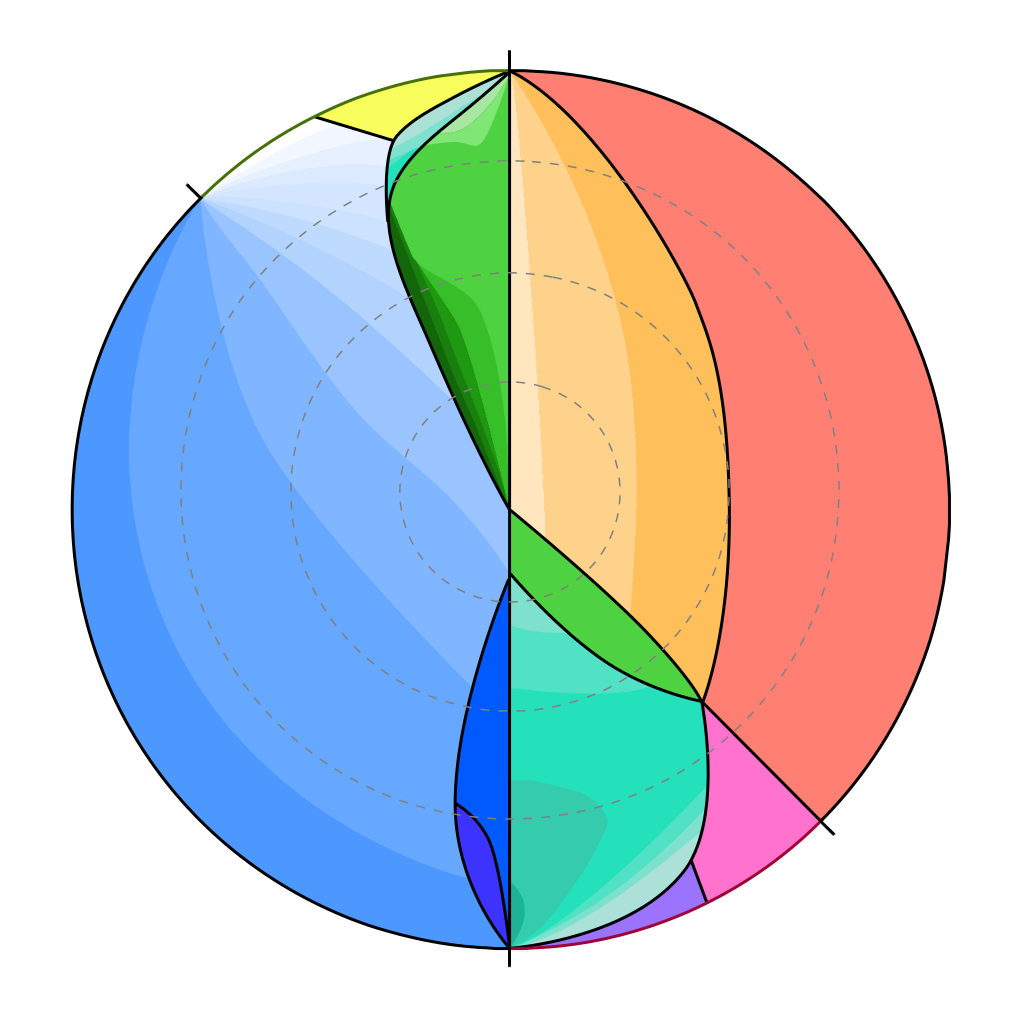}
        \put(68,150){FM$_a$}
        \put(191,150){AF$_c$}
        \put(130,150){AF-SZ$_{\text{AF}}$}
        \put(102,185){\ospa{-}}
        \put(130,108){\rotatebox[origin=c]{-40}{\ospa{+}}}
        \put(103,227){\color{black}\line(1,-5){3.3}}
        \put(102,202){\ospb{-}}
        \put(130,75){\ospb{+}}
        \put(130,39){SS$_{x\negmedspace/\negmedspace y}$}
        \put(150,42){\color{black}\line(2,-1){10}}
        \put(150,53){SS$_{b}$}
        \put(165,56){\color{black}\line(6,-1){15}}
        \put(83,46){MQ$_{1}$}
        \put(102,49){\color{black}\line(6,-1){15}}
        \put(81,76){MQ$_{2}$}
        \put(99,79){\color{black}\line(6,-1){15}}
        \put(66,202){SZ$_{x\negmedspace/\negmedspace y}$}
        \put(77,213){\color{black}\line(1,1){15}}
        \put(66,235){\rotatebox[origin=c]{20}{HK-SZ}}
        \put(212,185){\rotatebox[origin=c]{-60}{HK-AF}}
        \put(152,25){\rotatebox[origin=c]{20}{HK-SS}}
        \put(15,75){\rotatebox[origin=c]{-60}{HK-FM}}
        \put(128,249){\textcolor{darkgray}{\scriptsize{\textit{AF Kitaev}}}}
        \put(84,15){\textcolor{darkgray}{\scriptsize{\textit{FM Kitaev}}}}
        \put(215,134){\textcolor{darkgray}{\scriptsize{\textit{AF}}}}
        \put(191,126){\textcolor{darkgray}{\scriptsize{\textit{Heisenberg}}}}
        \put(21,134){\textcolor{darkgray}{\scriptsize{\textit{FM}}}}
        \put(21,126){\textcolor{darkgray}{\scriptsize{\textit{Heisenberg}}}}
        \put(119,250){\textcolor{darkgray}{\scriptsize{$\frac{\pi}{2}$}}}
        \put(35,215){\textcolor{darkgray}{\scriptsize{$\frac{3\pi}{4}$}}}
        \put(202,48){\textcolor{darkgray}{\scriptsize{$\frac{7\pi}{4}$}}}
        \put(118,15){\textcolor{darkgray}{\scriptsize{$\frac{3\pi}{2}$}}}
        \put(230,130){\textcolor{darkgray}{\scriptsize{$\phi=0$}}}
        \put(10,130){\textcolor{darkgray}{\scriptsize{$\pi$}}}
        \put(42,85){\textcolor{darkgray}{\scriptsize{$\theta=\frac{3\pi}{8}$}}}
        \put(79,101){\textcolor{darkgray}{\scriptsize{$\frac{\pi}{4}$}}}
        \put(102,117){\textcolor{darkgray}{\scriptsize{$\frac{\pi}{8}$}}}
      \end{overpic}}
  }

  \subfloat[][\har{} model]{
    \label{fig:pd_har_pos}
    \fbox{\begin{overpic}[scale=1,clip=true,trim=0 -10 0 0]{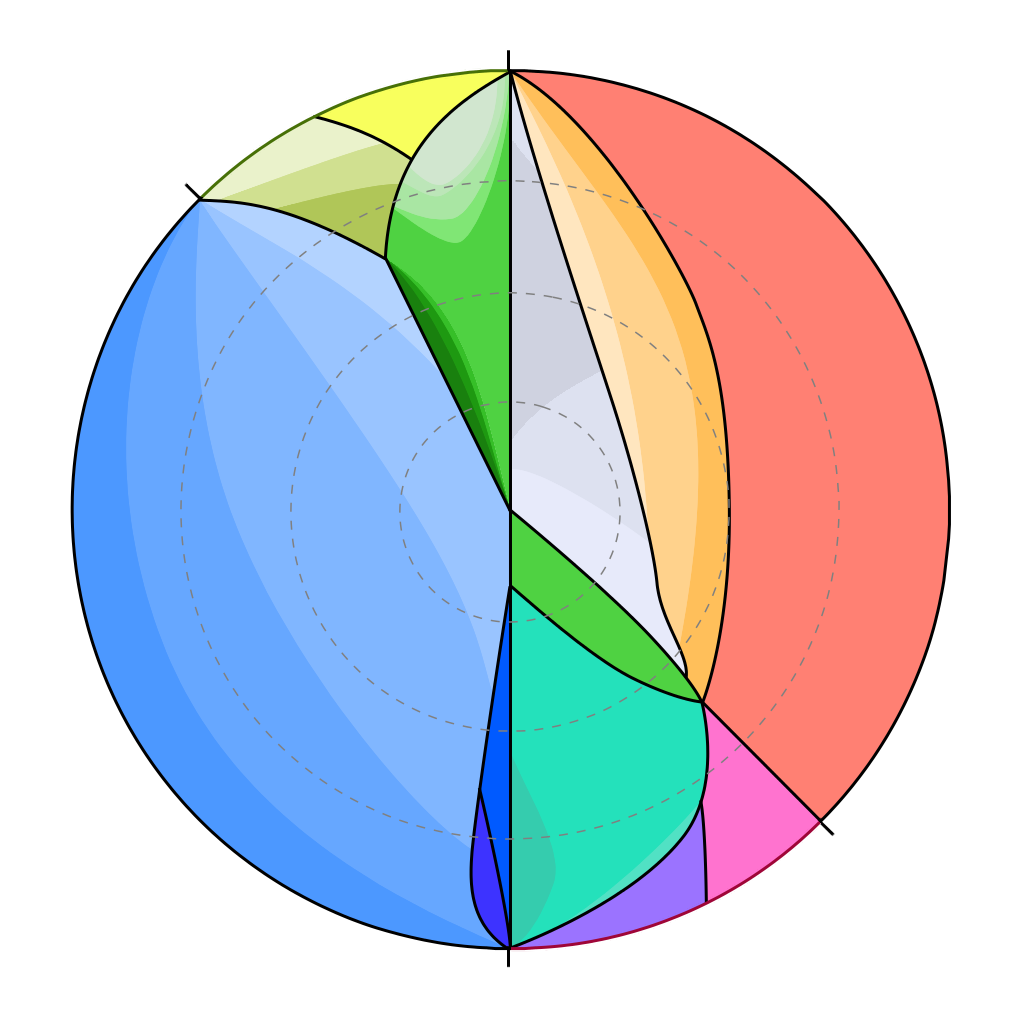}
        \put(68,150){FM$_a$}
        \put(191,150){AF$_c$}
        \put(147,170){AF-}
        \put(150,160){SZ$_{\text{AF}}$}
        \put(127,150){FM-}
        \put(130,140){SZ$_{\text{AF}}$}
        \put(102,185){\ospa{-}}
        \put(130,108){\rotatebox[origin=c]{-40}{\ospa{+}}}
        \put(105,233){\color{black}\line(1,-5){3}}
        \put(102,210){SZ$_{x\negmedspace/\negmedspace y}$}
        \put(130,75){\ospb{+}}
        \put(130,46){SS$_{x\negmedspace/\negmedspace y}$}
        \put(151,50){\color{black}\line(1,-1){10}}
        \put(148,60){SS$_{b}$}
        \put(163,63){\color{black}\line(6,-1){17}}
        \put(83,46){MQ$_{1}$}
        \put(102,49){\color{black}\line(6,-1){15}}
        \put(88,76){MQ$_{2}$}
        \put(106,79){\color{black}\line(6,-1){15}}
        \put(69,210){SZ$_{b}$}
        \put(66,235){\rotatebox[origin=c]{20}{HK-SZ}}
        \put(212,185){\rotatebox[origin=c]{-60}{HK-AF}}
        \put(152,25){\rotatebox[origin=c]{20}{HK-SS}}
        \put(15,75){\rotatebox[origin=c]{-60}{HK-FM}}
        \put(128,249){\textcolor{darkgray}{\scriptsize{\textit{AF Kitaev}}}}
        \put(84,15){\textcolor{darkgray}{\scriptsize{\textit{FM Kitaev}}}}
        \put(215,134){\textcolor{darkgray}{\scriptsize{\textit{AF}}}}
        \put(191,126){\textcolor{darkgray}{\scriptsize{\textit{Heisenberg}}}}
        \put(21,134){\textcolor{darkgray}{\scriptsize{\textit{FM}}}}
        \put(21,126){\textcolor{darkgray}{\scriptsize{\textit{Heisenberg}}}}
        \put(119,250){\textcolor{darkgray}{\scriptsize{$\frac{\pi}{2}$}}}
        \put(35,215){\textcolor{darkgray}{\scriptsize{$\frac{3\pi}{4}$}}}
        \put(202,48){\textcolor{darkgray}{\scriptsize{$\frac{7\pi}{4}$}}}
        \put(118,15){\textcolor{darkgray}{\scriptsize{$\frac{3\pi}{2}$}}}
        \put(230,130){\textcolor{darkgray}{\scriptsize{$\phi=0$}}}
        \put(10,130){\textcolor{darkgray}{\scriptsize{$\pi$}}}
        \put(42,85){\textcolor{darkgray}{\scriptsize{$\theta=\frac{3\pi}{8}$}}}
        \put(79,101){\textcolor{darkgray}{\scriptsize{$\frac{\pi}{4}$}}}
        \put(102,117){\textcolor{darkgray}{\scriptsize{$\frac{\pi}{8}$}}}
      \end{overpic}}
  }
  \caption{\label{fig:pds}(Color online) Classical phase diagrams for
    the $J\text{--}K\text{--}\Gamma$ pseudospin model with
    $\Gamma\ge0$.  The parametrization of the exchange interactions
    can be found in Eq. \ref{eq:parametrization}.  A detailed
    description of the phases can be found in Sec. \ref{sec:orders}
    while a summary can be found in Table \ref{tbl:orders}.  The color
    contours are guides for the eye: in the case of spiral (SP)
    states, they represent the length of the $\mathbf{Q}$-vector, whereas in
    the case of non-spiral states, they represent properties relevant
    to that particular phase; see Sec. \ref{sec:orders} for
    details.}
\end{figure}
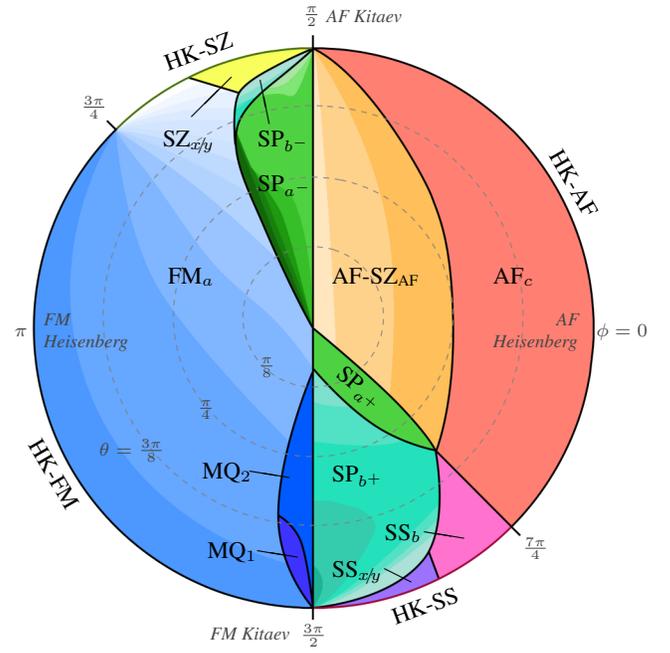
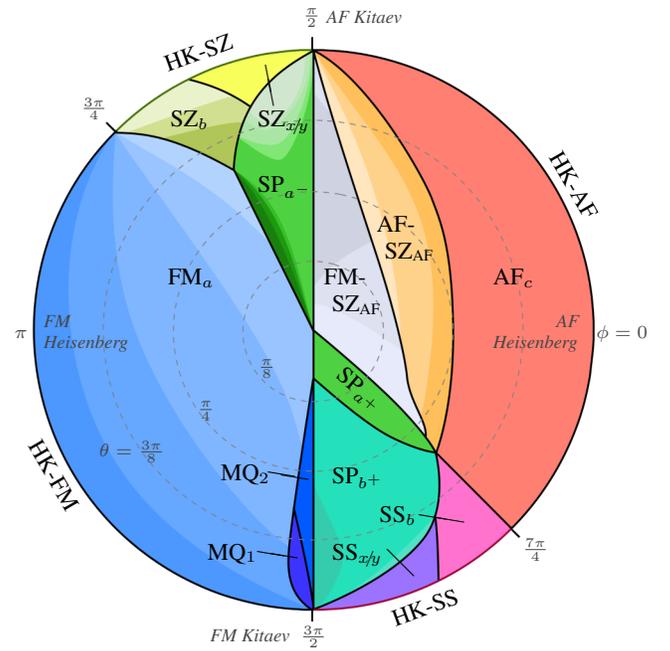

Before delving into the details of the classical magnetic ground
states of Eq. \ref{eq:ham}, we first examine some general features of
the \hhc{} and \har{} phase diagrams, which were obtained via the
methods outlined in Sec. \ref{subsec:methods}.

To fix the overall energy scale, we parametrize the $(J,K,\Gamma)$
parameter space using an angular representation
\begin{equation}
  \label{eq:parametrization}
  (J,K,\Gamma)=(\sin \theta \cos \phi, \sin \theta \sin \phi, \cos \theta),
\end{equation}
such that $\sqrt{J^2+K^2+\Gamma^2}=1$. The phase diagrams are plotted
as polar plots, where the angular component is given by $\phi$ and the
radial component, $r$, is given by $\theta$.  In
Figs. \ref{fig:pd_hhc_pos} and \ref{fig:pd_har_pos}, we show the phase
diagrams for the \hhc{} and \har{} lattices, respectively, when
$\Gamma\ge 0$ (\textit{i.e.}  $\phi\in[0,2\pi)$ and
$r=\theta\in[0,\frac{\pi}{2}]$); the $\Gamma\le 0$ results can be
obtained by applying time-reversal on the odd sublattice pseudospins,
which transforms $(J,K,\Gamma)\rightarrow(-J,-K,-\Gamma)$ and can be
seen in Fig. \ref{fig:pds_neg}.  Prior to Sec. \ref{subsec:mgamma}, we
will concentrate on the $\Gamma\ge0$ case with the understanding that
equivalent statements can be made for the $\Gamma\le0$ case.
Important properties of the $\Gamma\le 0$ phases---especially those
relating to experimental results---will be discussed in
Sec. \ref{subsec:mgamma}.

At first glance, we note the striking similarities between the two
phase diagrams: despite the different topology of the \hhc{} and
\har{} lattices, the parameter regimes where we find the various
magnetic orders and phase boundaries are similar in both systems.
This notable result emphasizes the commonalities shared between the
two systems as far as local physics is concerned.  On the other hand,
when compared to the classical phase diagram of the 2D honeycomb
iridate from Ref. \onlinecite{rau2014generic}, we notice that the
zigzag region has reduced in size and the spiral phases have become
more prevalent.  In particular, the 120$^\circ$ phase of the 2D model
is now a spiral phase in the two 3D models.

On the outer edges of the two phase diagrams, $\Gamma$ vanishes and
the Hamiltonians reach the isotropic HK limit.  With increasing
$\phi$, we encounter the N\'{e}el (HK-AF), skew-zigzag (HK-SZ),
ferromagnet (HK-FM), and skew-stripy phases (HK-SS), which were
discussed in the context of the 2D-honeycomb and 3D-honeycomb
iridates.\cite{chaloupka2010kitaev, reuther2011finite,
  price2013finite, lee2014heisenberg, lee2014order, kimchi2013three}
When $\Gamma=0$, the classical ground state manifold possesses an
emergent $SU(2)$ symmetry despite the presence of SOC.  However, both
classical and quantum order-by-disorder (ObD) lift this emergent
degeneracy and choose particular spatial directions for the
moments.\cite{chaloupka2010kitaev, price2013finite, lee2014heisenberg,
  lee2014order} Indeed, finite $\Gamma$ also breaks the emergent
symmetry\cite{rau2014generic} and pins pseudospin moments in
particular spatial directions, causing the phases on the inside of the
phase diagram to be non-collinear in general.

In the HK limit, the four-sublattice
transformation\cite{khaliullin2005orbital, chaloupka2010kitaev,
  kimchi2014kitaev} exactly maps the parameter point $(J,K)$ to
$(-J,K-2J)$ even in the quantum limit for both the
\hhc{}\cite{lee2014heisenberg, lee2014order} and
\har{}\cite{kimchi2013three} systems, thus relating the left and right
edges of the phase diagram. In the presence of finite $\Gamma$,
however, the four-sublattice transformation is not longer useful (it
does not map a parameter point with finite $\Gamma$ to another point
within the phase diagram); the left and right \textit{interiors} of
the phase diagrams are no longer related.

The centers of the phase diagrams are the $\Gamma$-only points.  These
points, much like in the 2D-honeycomb case, possess highly degenerate
ground state manifolds.\cite{rau2014generic} Therefore, similar to the
Kitaev points, small perturbations can drive different magnetic
orders.  As a result, we see that a number of phases converge at these
highly-degenerate points.

\subsection{Presence of new magnetic orders}

\begin{figure}[htbp!]
  \centering
  \setlength\fboxsep{0pt}
  \setlength\fboxrule{0pt}
  \subfloat[][\harzero{}: \ospa{+}]{
    \label{fig:hhc_spa_p}
    \fbox{\begin{overpic}[scale=1,clip=true,trim=0 0 0 0]{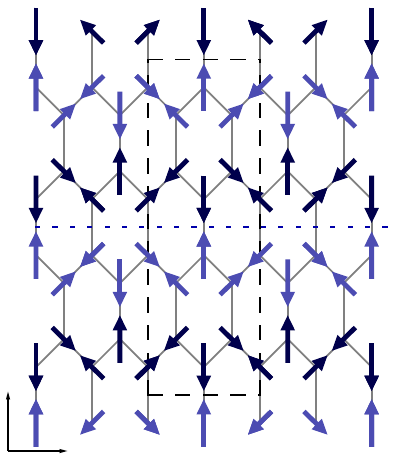}
        \put(0,21){\scriptsize{$c$}}
        \put(21,1){\scriptsize{$a$}}
        \put(-3,64){\scriptsize{$C^a_2$}}
      \end{overpic}}
  }
  \subfloat[][\har{}: \ospa{+}]{
    \label{fig:har_spa_p}
    \fbox{\begin{overpic}[scale=1,clip=true,trim=0 0 0 0]{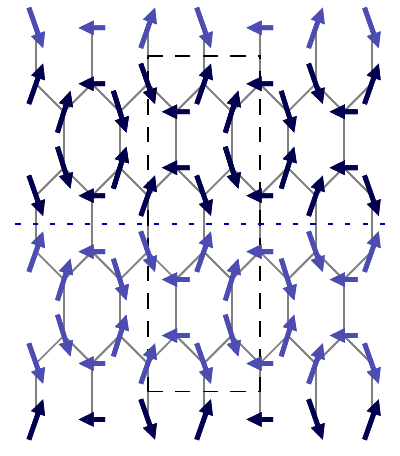}
      \end{overpic}}
  }

  \subfloat[][\harzero{}: \ospa{-}]{
    \label{fig:hhc_spa_m}
    \fbox{\begin{overpic}[scale=1,clip=true,trim=0 0 0 0]{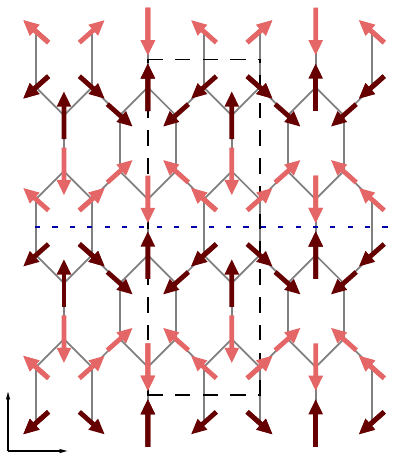}
        \put(0,21){\scriptsize{$c$}}
        \put(21,1){\scriptsize{$a$}}
        \put(-3,64){\scriptsize{$C^a_2$}}
      \end{overpic}}
  }
  \subfloat[][\har{}: \ospa{-}]{
    \label{fig:har_spa_m}
    \fbox{\begin{overpic}[scale=1,clip=true,trim=0 0 0 0]{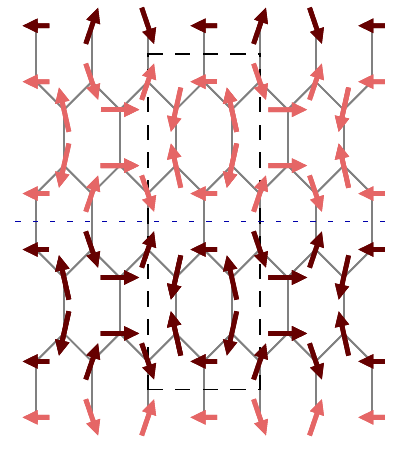}
      \end{overpic}}
  }
  \caption{\label{fig:realspace_a}(Color online) Real-space spin
    configurations of the \ospa{} spiral states obtained from
    simulated annealing with spins projected on to the $ac$-plane.
    The chosen parameter points yield $\mathbf{Q}=(0.33,0,0)$;
    however, we note that the $\mathbf{Q}$-vector in these phases are
    generally incommensurate and are not the same as the peak
    positions of the structure factor (see main text and
    Figs. \ref{fig:sf_a}, \ref{fig:sf_b}).  The black dashed boxes
    enclose the conventional unit cells.  Identical colors indicate
    that the sublattices share spiral-planes.  Shades of blue indicate
    that the spiral-planes are aligned with the honeycomb-planes while
    shades of red indicate that the spiral-planes are not aligned with
    the honeycomb-planes.  The handedness of adjacent sites can be
    readily verified as being opposite: the spirals counter-propagate.
    Examples of the preserved rotation symmetry are indicated by the
    dotted blue lines.}
\end{figure}

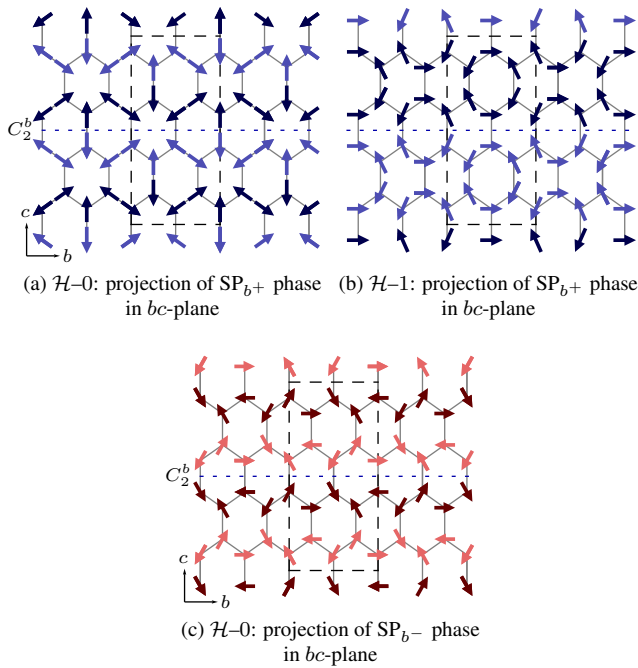
\begin{figure}[htbp!]
  \centering
  \setlength\fboxsep{0pt}
  \setlength\fboxrule{0pt}
  \subfloat[][\harzero{}: projection of \ospb{+} phase in $bc$-plane]{
    \label{fig:hhc_spb_p}
    \fbox{\begin{overpic}[scale=1,clip=true,trim=0 0 0 0]{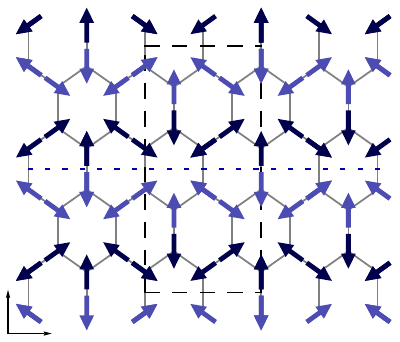}
        \put(0,17){\scriptsize{$c$}}
        \put(16,0){\scriptsize{$b$}}
        \put(-5,48){\scriptsize{$C^b_2$}}
      \end{overpic}}
  }
  \subfloat[][\har{}: projection of \ospb{+} phase in $bc$-plane]{
    \label{fig:har_spb_p}
    \fbox{\begin{overpic}[scale=1,clip=true,trim=0 0 0 0]{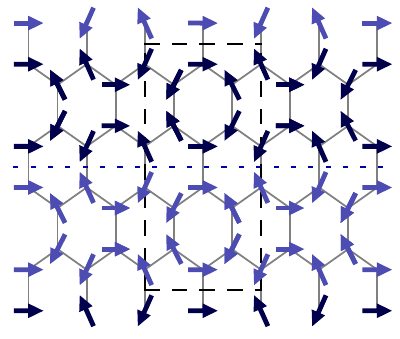}
      \end{overpic}}
  }

  \subfloat[][\harzero{}: projection of \ospb{-} phase in $bc$-plane]{
    \label{fig:hhc_spb_m}
    \fbox{\begin{overpic}[scale=1,clip=true,trim=0 0 0 0]{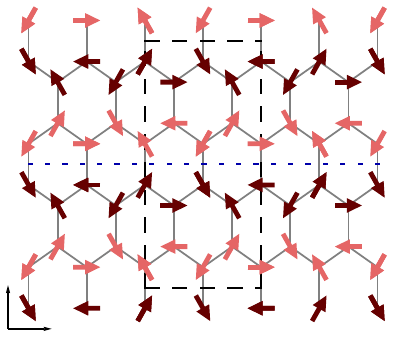}
        \put(0,17){\scriptsize{$c$}}
        \put(16,0){\scriptsize{$b$}}
        \put(-5,48){\scriptsize{$C^b_2$}}
      \end{overpic}}
  }
  \caption{\label{fig:realspace_b}(Color online) Real-space spin
    configurations of the \ospb{} spiral states obtained from
    simulated annealing with spins projected on to the $bc$-plane.
    The chosen parameter points yield $\mathbf{Q}=(0,0.33,0)$;
    however, we note that the $\mathbf{Q}$-vector in these phases are
    generally incommensurate.  See caption in
    Fig. \ref{fig:realspace_a} for further details.  In the case of
    the \ospb{-} phase, only the spiral planes and not the spins
    preserve the $C^b_2$ rotation symmetry.}
\end{figure}

The effects of a finite $\Gamma$ is most pronounced near the Kitaev
points where the highly degenerate classical ground state manifolds
can be lifted by small perturbations.  As a result, finite $\Gamma$ in
these regions generates new magnetic orders that are not continuously
connected with those found along the HK limit.  Of the eight new
magnetic orders, four are incommensurate spiral phases (\ospa{-},
\ospa{+}, \ospb{+}, and \ospb{-}), two are multiple-$\mathbf{Q}$ states (MQ$_1$
and MQ$_2$), one is an antiferromagnetic state (AF-SZ$_{\text{AF}}$),
and one has finite net moment (FM-SZ$_{\text{AF}}$).  Of the four
spiral states, two possess $\mathbf{Q}$-vectors that vary in length depending
on the ratio of the exchange parameters while all of them contain
counter-propagating spiral patterns (see Sec. \ref{sec:orders} for
details).

\subsection{Magnetic orders continuously connected to the
  Heisenberg-Kitaev limit}
Away from the Kitaev points, finite but small $\Gamma$ modifies
existing magnetic orders found in the HK limit.  Several of the
existing phases---namely the HK-SZ, HK-FM, and HK-SS
phases---continuously deform with increasing $\Gamma$ to acquire
additional spin textures.  The details of these additional spin
textures depend on the particular lattice in consideration, which we
will outline in Sec. \ref{sec:orders}.

\section{\label{sec:orders}Description of the magnetic orders}
In this section, we provide detailed descriptions of the magnetic
phases found in our phase diagrams.  We first discuss the new magnetic
orders, then we will describe the phases that are continuously
connected with those at the HK limit.  We also present the static
structure factors of the spiral phases, which are relevant to x-ray
and neutron scattering experiments.  A summary of the phases discussed
in this section can be found in Table \ref{tbl:orders}.

\subsection{Emergent magnetic orders}
\paragraph*{Spiral orders:}
The spiral phases are those that can be represented by the ansatz in
Eq. \ref{eq:singleQAnsatz} with incommensurate $\mathbf{Q}$-vectors.  They are
located where the Kitaev exchange dominates over the Heisenberg
exchange and where the two exchanges compete and have opposite signs.
These spiral regions are connected to both the Kitaev and the
pure-$\Gamma$ points and border the $J=0$ line.  The above description
of the location of spirals states is valid for both $\Gamma\ge0$ and
$\Gamma\le0$.

There are three spiral phases shared by both lattices: \ospa{+},
\ospa{-}, and \ospb{+}.  In addition, \ospb{-} is found in the \hhc{}
lattice in a small region near the \ospa{-} phase.  All these spiral
phases do not have a net moment and they are all non-coplanar.

In the FM Heisenberg-AF Kitaev region (quadrant II of
Fig. \ref{fig:pds}), we find the \ospa{-} phase.  This phase has a
$\mathbf{Q}$-vector of the form $(h00)$ (in the orthorhombic hkl
notation), signifying that the spirals run along the $\hat{a}$
direction; therefore, we label this state as \ospa{-}.  In the \hhc{}
lattice, there is also the nearby \ospb{-} phase, which has a
$\mathbf{Q}$-vector of the form $(0k0)$.

In the AF Heisenberg-FM Kitaev region (quadrant IV), there are two
spiral phases: the smaller region has a $\mathbf{Q}$-vector of the
form $(h00)$ while the larger region has $(0k0)$.  We label the former
\ospa{+} and the latter \ospb{+}.

We illustrate an example of the real-space spin configuration for
these phases in Figs. \ref{fig:realspace_a} and \ref{fig:realspace_b}
using relevant projections that highlight the spin spirals.  For the
\ospa{} phases, the spins have been projected on to the $ac$ plane
while, for the \ospb{} phases, the projection is on to the $bc$
plane. Although we have chosen parameter points that give commensurate
$\mathbf{Q}$-vectors, we note that generically the phases are
incommensurate spiral phases.

Prior to expounding the features of these spiral states, we emphasize
that the quoted $\mathbf{Q}$-vectors are those that would appear in
the single-$\mathbf{Q}$ ansatz of Eq. \ref{eq:singleQAnsatz}, which
are defined up to reciprocal lattice vectors of the conventional
Brillouin zone.  We stress that these $\mathbf{Q}$-vectors are
related, but not necessarily the same as, the peak positions of the
static structure factor, which we denote by $\mathbf{q}$.  The static
structure factor is given by
\begin{align}
  \mathcal{S}(\mathbf{q})&=\frac{1}{N}\sum_{ij}\langle\mathbf{S}_i\cdot \mathbf{S}_j\rangle e^{-i \mathbf{q}\cdot (\mathbf{r}_i-\mathbf{r}_j)} \nonumber \\
  &=\frac{1}{N}\sum_{ab}\langle\mathbf{s}^{ab}_{\mathbf{q}} \rangle
  e^{-i \mathbf{q}\cdot \mathbf{r}_{ab}},
\end{align}
where $\mathbf{S}_i$ is the spin at site $i$ of the ground state
configuration and $\mathbf{r}_i$ is its real-space position.  In the
second line, we summed over the unit cells but retained the sublattice
indices such that the static structure factor is written in terms of
the matrix spin-spin structure factor
$\mathbf{s}^{ab}_{\mathbf{q}}=\mathbf{S}^a_{\mathbf{q}}\cdot\mathbf{S}^b_{\mathbf{-q}}$
and relative sublattice positions
$\mathbf{r}_{ab}=\mathbf{r}_a-\mathbf{r}_b$.  In this second form, we
see explicitly that single-$\mathbf{Q}$ ansatz can have peaks at
$\mathbf{q}=\mathbf{g}\pm\mathbf{Q}$ where $\mathbf{g}$ is a
reciprocal lattice vector of the conventional Brillouin zone (one of
the form $(hkl)$ where all indices are integers).  However, because of
the form factor $e^{-i\mathbf{q}\cdot\mathbf{r}_{ab}}$ and the matrix structure of
$\mathbf{s}^{ab}_{\mathbf{q}}$, not all peaks will have the same
intensity and some may even be extinguished.

In Figs. \ref{fig:sf_a} and \ref{fig:sf_b}, we show the static
structure factors along relevant cuts in reciprocal space which
highlights the difference between $\mathbf{Q}$ and $\mathbf{q}$.
Although the shown \ospa{} phases have the same
$\mathbf{Q}=(0.33,0,0)$, \ospa{-} possesses a peak at
$\mathbf{q}=\left[(100)-\mathbf{Q}\right]=(0.66,0,0)$ while \ospa{+} does not.
Instead, \ospa{+} possesses a peak at
$\mathbf{q}=\left[(100)+\mathbf{Q}\right]=(1.33,0,0)$.  Likewise for the \ospb{}
phases: \ospb{-} and \ospb{+} differ in that \ospb{-} possesses a peak
at $\mathbf{q}=\left[(010)-\mathbf{Q}\right]$ while the closest peak along the
$(0k0)$ direction for \ospb{+} is located at
$\mathbf{q}=\left[(010)+\mathbf{Q}\right]$.  This property is generic for all
$\mathbf{Q}$-vectors encountered in the spiral phases and hence
explain our chosen notation: the spiral phase SP$_{x\pm}$ possesses
the lowest-order structure factor peak at
$\mathbf{q}=(\mathbf{1}_x\pm\mathbf{Q})$, where $\mathbf{1}_x$ is
short form for $(100)$ or $(010)$ depending on whether $x$ is $a$ or
$b$.  This result is correlated with the orientation of the
\textit{spiral-planes} of these phases, which we will address after
introducing several properties of these spiral-planes.

Although these are non-coplanar spirals, spins on any given sublattice
all lie in the same plane, which we term \textit{spiral-plane} of the
sublattice/site (this may not be evident in
Figs. \ref{fig:realspace_a} and \ref{fig:realspace_b} due to the
projections).  In other words, the non-coplanar nature of these spiral
phases is due to different sublattices having different spiral-planes.
Specifically, sublattices that share honeycomb-planes also share
spiral-planes,\footnote{In the \har{}, the spiral-planes amongst sites
  in the honeycomb strips are slightly askew: bridging-sites share
  spiral-planes as do non-bridging-sites, but these two
  symmetry-inequivalent sites have spiral-planes that are slightly
  off-angle relative to each other (in a manner where the $c$-axis
  still lies in the spiral-planes).}  whereas sublattices that don't
share honeycomb-planes have spiral-planes related by symmetry: in the
\ospa{} phases, both the spiral-planes and spins preserve the $C_{2}$
rotation symmetry about the $\hat{a}$ direction (hereby denoted as
$C^{a}_{2}$), while in the \ospb{+} phase, both the spiral planes and
spins preserve the $C^{b}_{2}$ rotation symmetry.  In \ospb{-}, the
spins break all $C_2$ symmetries but the spiral-planes are related by
the $C^{b}_{2}$ rotations.  In Figs. \ref{fig:realspace_a} and
\ref{fig:realspace_b}, sites that have the same spiral-planes have the
same color.  We have also illustrated the preserved symmetry
operations.

Despite possessing the same $\mathbf{Q}$-vector for each sublattice as evident
in Eq. \ref{eq:singleQAnsatz}, NNs that share spiral-planes have
spirals that propagate with different handedness, which can be readily
verified in Figs. \ref{fig:realspace_a} and \ref{fig:realspace_b}.  In
other words, the angles formed by NN spins that share spiral-planes
are not constant as the spiral propagates---we say that the two
spirals \textit{counter-propagate}.  We note that the corresponding
spiral phases in the $\Gamma\le0$ regime also possess
counter-propagating spirals.  This is because the time-reversal
operation on odd sublattices leaves the handedness of all spirals
invariant.

The orientations of the spiral-planes are related to the honeycomb
planes and distinguish the $+$ from the $-$ phases.  They are also
correlated with the positions of the peaks in the structure factors.
In both the \ospb{+} and \ospa{+} phase, the spiral-planes at each
site is aligned with the honeycomb-plane of that site.  On the other
hand, in the \ospa{-} and \ospb{-} phases, the spiral-planes at each
site is aligned with the other honeycomb-plane of the
lattice.\footnote{More precisely, the spiral-plane tilts slightly away
  from its related honeycomb-plane (in a manner where the $c$-axis
  still lies in the spiral-planes).  The angle between a spiral-plane
  and its related honeycomb-plane is a monotonic function of
  $\varphi=\tan^{-1}(K/J)\in(-\pi,\pi)$: in the \ospa{-} phase, the
  angle is approximately $0$ when $\varphi$ is near $\pi/2$ and
  increases as $\varphi$ moves towards $\pi$.  In the \ospb{+} phase,
  the angle increases from approximately $0$ when $\varphi$ is closest
  to $0$ and increases as $\varphi$ decreases towards $-\pi/2$.} This
implies that states that have structure factor peaks at
$\mathbf{q}=(\mathbf{1}_x\pm\mathbf{Q})$ have honeycomb-planes
aligned/not aligned with spiral planes.  In Fig. \ref{fig:realspace_a}
and \ref{fig:realspace_b}, we illustrate the alignment of the
spiral-planes and the honeycomb-planes by the hue of the sites: shades
of blue indicates that the site's spiral-plane and honeycomb-plane are
in alignment, shades of red indicate otherwise.

The lengths of the $\mathbf{Q}$-vectors are illustrated as a guide for the eye by
color contours in Fig. \ref{fig:pds}: the $\mathbf{Q}$-vectors lengthen from the
lightest to the darkest of colors within each phase.  In the \ospa{+}
and \ospb{-} phases, the $\mathbf{Q}$-vectors are approximately constant at
$(0.33,0,0)$ and $(0,0.33,0)$, respectively.  On the other hand, the
$\mathbf{Q}$-vector varies continuously within the \ospa{-} and \ospb{+} phases.

In the \ospa{-} phase of the \hhc{} lattice, the $\mathbf{Q}$-vector varies
between approximately $(0.20,0,0)$ and $(0.47,0,0)$ while in the
corresponding phase of the \har{} lattice, the $\mathbf{Q}$-vector varies
between approximately $(0.15,0,0)$ and $(0.43,0,0)$.  The majority of
the parameter space within the \ospa{-} of both lattices has
$\mathbf{Q}=(0.33,0,0)$, which are the states shown in
Fig. \ref{fig:realspace_a}.  In both lattices, the
$\mathbf{Q}=(0.33,0,0)$ states are located centrally within the phase
and border the $J=0$ line.  The longer $\mathbf{Q}$-vectors are found in a
narrow region bordering the FM$_a$ phase while the shorter $\mathbf{Q}$-vectors
occupy a small region towards the AF Kitaev point.

In the \ospb{+} of the \hhc{} lattice, the $\mathbf{Q}$-vector ranges
approximately from $(0,0.17,0)$ to $(0,0.40,0)$ while for the
corresponding phase in the \har{} lattice, the $\mathbf{Q}$-vector range is
smaller: approximately $(0,0.30,0)$ to $(0,0.36,0)$.  For this phase
in both lattices, the majority of the parameter space possesses the
$\mathbf{Q}$-vector $(0,0.33,0)$ and these are the states shown in
Fig. \ref{fig:realspace_b}.  The $(0,0.33,0)$ states are located
centrally within the phase.  The longer $\mathbf{Q}$-vectors are localized near
the FM Kitaev point, while the shorter $\mathbf{Q}$-vectors border the
SS$_{x/y}$ phase.

\paragraph*{MQ$_1$:}In quadrant III of the phase diagram where both
Heisenberg and Kitaev exchanges are ferromagnetic, there are two
multiple-$\mathbf{Q}$ states that cannot be expressed as a
single-$\mathbf{Q}$ ansatz given by Eq. \ref{eq:singleQAnsatz}.  The
first of which is a non-coplanar state near the FM Kitaev point and
borders both the FM$_a$ and \ospb{+} phases.  It has significant
Fourier components in $\mathbf{Q}=(0.5,0.5,x)$ for multiple values of
$x$ and does not have a finite net moment.

\paragraph*{MQ$_2$:}The second of the two multiple-$\mathbf{Q}$ states
lies closer to the pure-$\Gamma$ point than the first
multiple-$\mathbf{Q}$ state.  It also borders both the FM$_a$ and the
\ospb{+} phases and is a non-coplanar state.  This phase has
significant Fourier components in $\mathbf{Q}=(\pm0.33,\pm0.33,0)$,
$(\pm0.33,0,0)$, as well as $(0,0,0)$, the last of which indicates
that it possesses a finite net moment, which lies in the $\hat{a}$
direction.  A more detailed analysis of these multiple-$\mathbf{Q}$
states described here are left for future study.

\paragraph*{AF-SZ$_{AF}$:}Bordering the AF$_c$ phase is this coplanar
phase with vanishing net moment, which is composed of
antiferromagnetic chains along the skew-zigzag directions.  In the
case of the \hhc{} lattice, this phase can be verified via LTA as the
exact ground state.  This phase is connected to the AF Kitaev point
and also partially borders the \ospa{+} phase.  In both lattices, the
projections of the spins along the $\hat{c}$ direction vanishes.  In
the \hhc{} lattice, the AF chains along the skew-zigzag directions
are collinear and are nearly parallel to the honeycomb-plane of the
skew-zigzag.  The angle of deviation between the collinear spins and
the honeycomb-plane increases as one moves aways from the AF$_c$
phase.  This trend can be captured by the magnitude of the projection
of the spins in the $\hat{a}$, which is depicted as colour contours in
Fig. \ref{fig:pds}: the larger the projection along $\hat{a}$, the
darker the color.  In the \har{} lattice, the spins along the
skew-zigzag directions are not perfectly collinear, but nevertheless
the magnitude of the spin projections in the $\hat{a}$ direction also
follows the same trend.  Furthermore, the non-bridging-$z$-bonds are
antiferromagnetically correlated.  In both lattices, the phase
preserves the $C^a_2$ rotation symmetry.

\paragraph*{FM-SZ$_{AF}$:}This coplanar phase only exists in the
\har{} lattice and is the version of the AF-SZ$_{AF}$ phase with a
finite net moment.  This phase is bordered by the \ospa{-}, \ospa{+},
and AF-SZ$_{AF}$ phases in the AF-Heisenberg regime and is connected
to both the AF Kitaev and pure-$\Gamma$ points.  Like the AF-SZ$_{AF}$
phase of the \har{} lattice, spins have vanishing projections along
the $\hat{c}$ direction, they form near-collinear AF chains along the
zigzag directions, and the $C^a_2$ rotation symmetry is preserved.
These AF chains become progressively less collinear as one approaches
the $J=0$ line.

Unlike the AF-SZ$_{AF}$ phase, however, the non-bridging-$z$-bonds are
ferromagnetically correlated.  As a result, despite being in the
AF-Heisenberg regime, there is a net moment in the $\hat{a}$
direction.  The size of the net moment is small, especially compared
to the FM$_a$ phase, and decreases as we approach the \ospa{+} phase
or the AF Kitaev point.  We illustrate this decrease in net moment by
the lightening of the color contours.

\subsection{Existing magnetic orders}

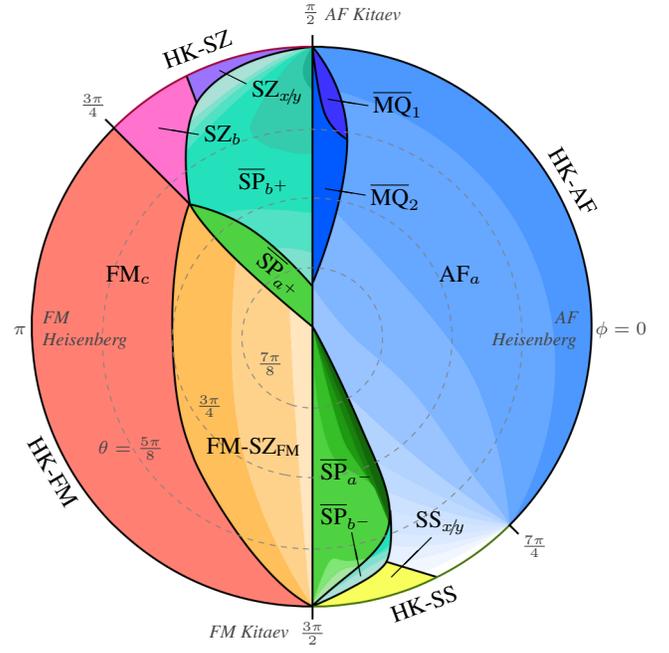
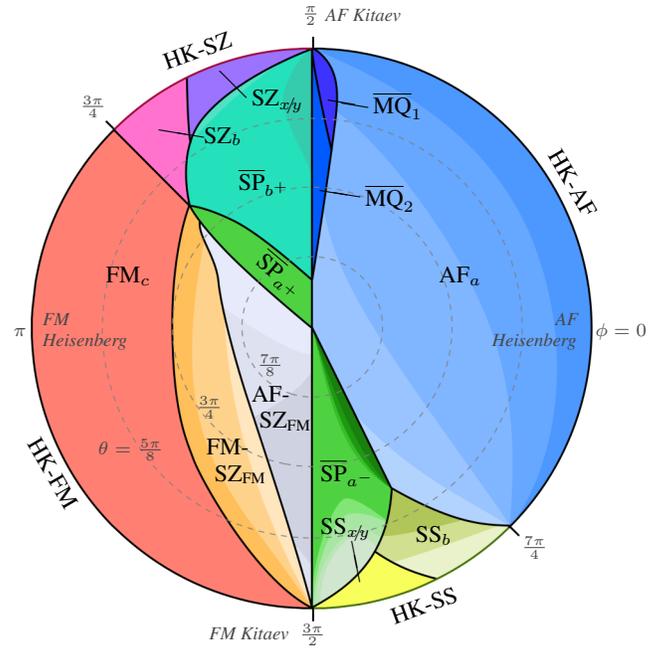
\begin{figure}[htbp!]
  \centering
  \setlength\fboxsep{0pt}
  \setlength\fboxrule{0pt}
  \subfloat[][Hyperhoneycomb model]{
    \label{fig:pd_hhc_pos}
    \fbox{\begin{overpic}[scale=1,clip=true,trim=0 -10 0 0]{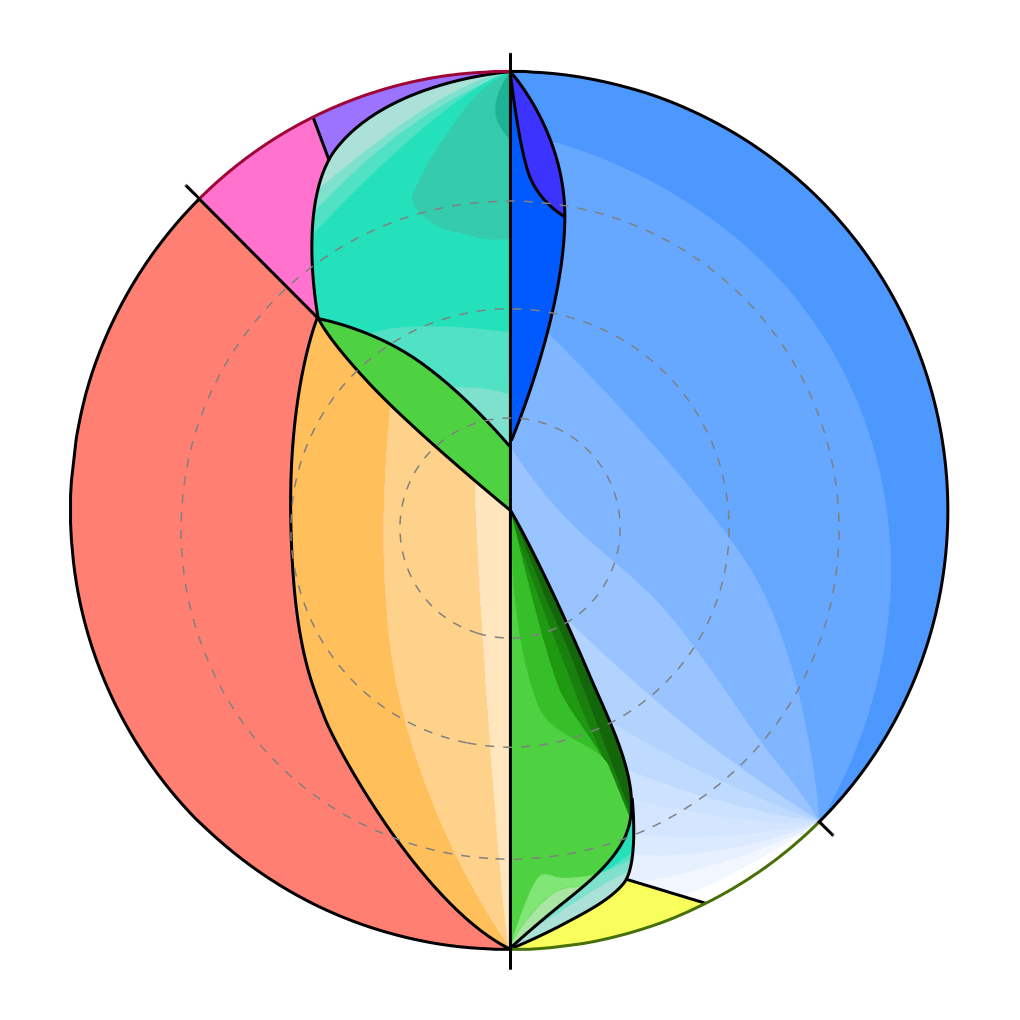}
        \put(45,150){FM$_c$}
        \put(171,150){AF$_a$}
        \put(83,85){FM-SZ$_{\text{FM}}$}
        \put(126,75){\ospabar{-}}
        \put(100,150){\rotatebox[origin=c]{-40}{\ospabar{+}}}
        \put(138,55){\color{black}\line(1,-5){3.3}}
        \put(126,58){\ospbbar{-}}
        \put(95,185){\ospbbar{+}}
        \put(100,220){SZ$_{x\negmedspace/\negmedspace y}$}
        \put(88,230){\color{black}\line(2,-1){10}}
        \put(82,203){SZ$_{b}$}
        \put(65,208){\color{black}\line(6,-1){15}}
        \put(146,214){$\overline{\text{MQ}}_{1}$}
        \put(129,219){\color{black}\line(6,-1){15}}
        \put(145,179){$\overline{\text{MQ}}_{2}$}
        \put(128,185){\color{black}\line(6,-1){15}}
        \put(162,57){SS$_{x\negmedspace/\negmedspace y}$}
        \put(153,38){\color{black}\line(1,1){15}}
        \put(66,235){\rotatebox[origin=c]{20}{HK-SZ}}
        \put(212,185){\rotatebox[origin=c]{-60}{HK-AF}}
        \put(152,25){\rotatebox[origin=c]{20}{HK-SS}}
        \put(15,75){\rotatebox[origin=c]{-60}{HK-FM}}
        \put(128,249){\textcolor{darkgray}{\scriptsize{\textit{AF Kitaev}}}}
        \put(84,15){\textcolor{darkgray}{\scriptsize{\textit{FM Kitaev}}}}
        \put(215,134){\textcolor{darkgray}{\scriptsize{\textit{AF}}}}
        \put(191,126){\textcolor{darkgray}{\scriptsize{\textit{Heisenberg}}}}
        \put(21,134){\textcolor{darkgray}{\scriptsize{\textit{FM}}}}
        \put(21,126){\textcolor{darkgray}{\scriptsize{\textit{Heisenberg}}}}
        \put(119,250){\textcolor{darkgray}{\scriptsize{$\frac{\pi}{2}$}}}
        \put(35,215){\textcolor{darkgray}{\scriptsize{$\frac{3\pi}{4}$}}}
        \put(202,48){\textcolor{darkgray}{\scriptsize{$\frac{7\pi}{4}$}}}
        \put(118,15){\textcolor{darkgray}{\scriptsize{$\frac{3\pi}{2}$}}}
        \put(230,130){\textcolor{darkgray}{\scriptsize{$\phi=0$}}}
        \put(10,130){\textcolor{darkgray}{\scriptsize{$\pi$}}}
        \put(42,85){\textcolor{darkgray}{\scriptsize{$\theta=\frac{5\pi}{8}$}}}
        \put(79,101){\textcolor{darkgray}{\scriptsize{$\frac{3\pi}{4}$}}}
        \put(102,117){\textcolor{darkgray}{\scriptsize{$\frac{7\pi}{8}$}}}
      \end{overpic}}
  }

  \subfloat[][\har{} model]{
    \label{fig:pd_har_pos}
    \fbox{\begin{overpic}[scale=1,clip=true,trim=0 -10 0 0]{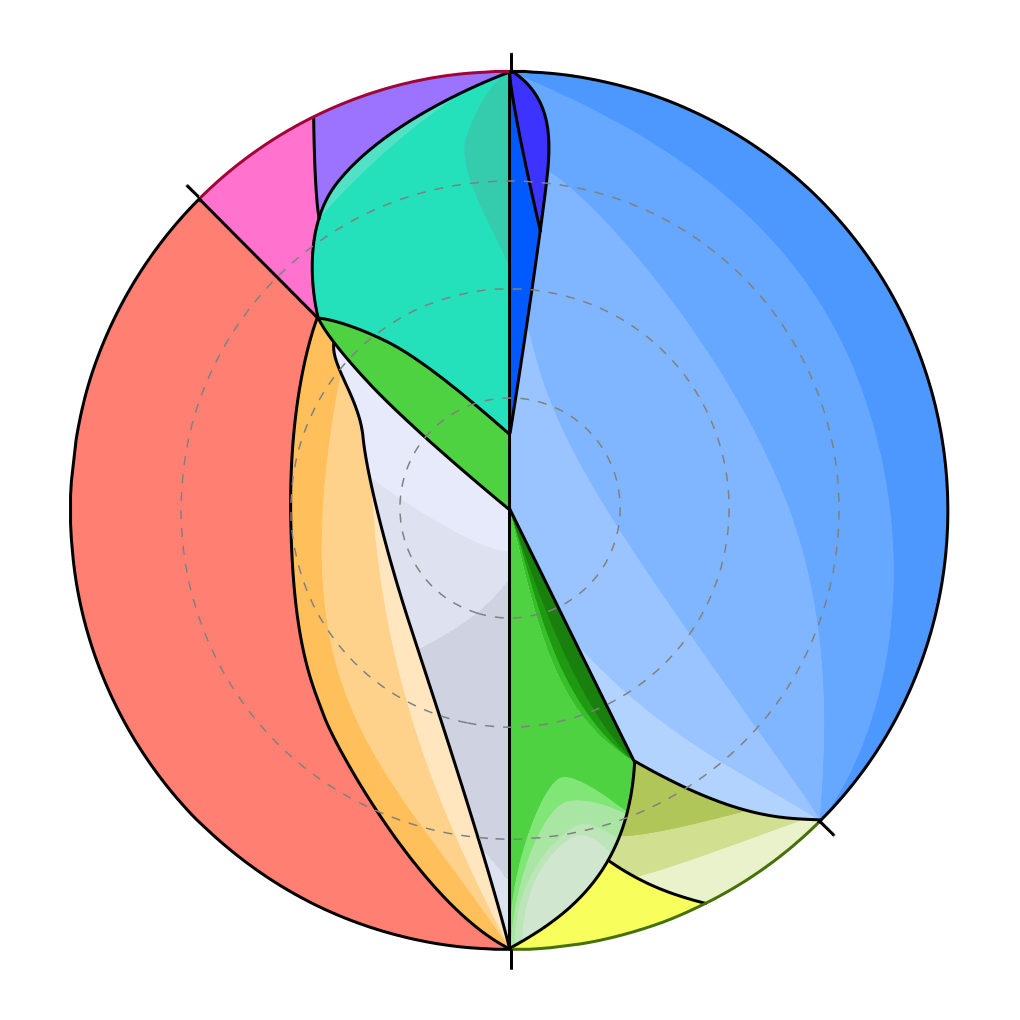}
        \put(45,150){FM$_c$}
        \put(171,150){AF$_a$}
        \put(83,85){FM-}
        \put(86,75){SZ$_{\text{FM}}$}
        \put(100,105){AF-}
        \put(103,95){SZ$_{\text{FM}}$}
        \put(126,75){\ospabar{-}}
        \put(100,150){\rotatebox[origin=c]{-40}{\ospabar{+}}}
        \put(138,52){\color{black}\line(1,-5){3}}
        \put(126,55){SS$_{x\negmedspace/\negmedspace y}$}
        \put(95,185){\ospbbar{+}}
        \put(100,217){SZ$_{x\negmedspace/\negmedspace y}$}
        \put(88,230){\color{black}\line(1,-1){10}}
        \put(82,203){SZ$_{b}$}
        \put(65,208){\color{black}\line(6,-1){17}}
        \put(146,214){$\overline{\text{MQ}}_{1}$}
        \put(129,219){\color{black}\line(6,-1){15}}
        \put(143,179){$\overline{\text{MQ}}_{2}$}
        \put(126,185){\color{black}\line(6,-1){15}}
        \put(162,52){SS$_{b}$}
        \put(66,235){\rotatebox[origin=c]{20}{HK-SZ}}
        \put(212,185){\rotatebox[origin=c]{-60}{HK-AF}}
        \put(152,25){\rotatebox[origin=c]{20}{HK-SS}}
        \put(15,75){\rotatebox[origin=c]{-60}{HK-FM}}
        \put(128,249){\textcolor{darkgray}{\scriptsize{\textit{AF Kitaev}}}}
        \put(84,15){\textcolor{darkgray}{\scriptsize{\textit{FM Kitaev}}}}
        \put(215,134){\textcolor{darkgray}{\scriptsize{\textit{AF}}}}
        \put(191,126){\textcolor{darkgray}{\scriptsize{\textit{Heisenberg}}}}
        \put(21,134){\textcolor{darkgray}{\scriptsize{\textit{FM}}}}
        \put(21,126){\textcolor{darkgray}{\scriptsize{\textit{Heisenberg}}}}
        \put(119,250){\textcolor{darkgray}{\scriptsize{$\frac{\pi}{2}$}}}
        \put(35,215){\textcolor{darkgray}{\scriptsize{$\frac{3\pi}{4}$}}}
        \put(202,48){\textcolor{darkgray}{\scriptsize{$\frac{7\pi}{4}$}}}
        \put(118,15){\textcolor{darkgray}{\scriptsize{$\frac{3\pi}{2}$}}}
        \put(230,130){\textcolor{darkgray}{\scriptsize{$\phi=0$}}}
        \put(10,130){\textcolor{darkgray}{\scriptsize{$\pi$}}}
        \put(42,85){\textcolor{darkgray}{\scriptsize{$\theta=\frac{5\pi}{8}$}}}
        \put(79,101){\textcolor{darkgray}{\scriptsize{$\frac{3\pi}{4}$}}}
        \put(102,117){\textcolor{darkgray}{\scriptsize{$\frac{7\pi}{8}$}}}
      \end{overpic}}
  }
  \caption{\label{fig:pds_neg}(Color online) Classical phase diagrams
    for the $J\text{--}K\text{--}\Gamma$ pseudospin model with
    $\Gamma\le0$.  The details of this phase diagram can be understood
    via a classical mapping that relates
    $(J,K,\Gamma)\rightarrow(-J,-K,-\Gamma)$; see
    Sec. \ref{subsec:mgamma} for details.  The color contours are
    guides for the eye: in the case of spiral (SP) states, they
    represent the length of the $\mathbf{Q}$-vector, whereas in the
    case of non-spiral states, they represent properties relevant to
    that particular phase; see Sec. \ref{sec:orders} for details.}
\end{figure}

\paragraph*{AF$_{c}$:}This collinear phase is the exact ground state
in the AF-Heisenberg region when $J>-K$ and $J>\Gamma$: the LTA
succeeds in finding this exact ground state for both lattices.  This
phase is the HK-AF state with moments aligned antiferromagnetically
and locked in the $\hat{c}$ direction due to the presence of $\Gamma$.

\paragraph*{FM$_{a}$:}This coplanar state with finite net moment
encompasses a large fraction of both phase diagrams.  Only in the
\hhc{} does the LTA succeed in identifying this phase as the exact
ground state.  In both lattices, the projection of the spins along the
$\hat{a}$ direction is ferromagnetic while the projection along the
$\hat{c}$ direction vanishes.  The projection along the $\hat{b}$
direction behaves differently for the \hhc{} and the \har{} models.
In the \hhc{} lattice, the $\hat{b}$ component orders in the
skew-zigzag order, while in the \har{} lattice, the $\hat{b}$
component is ferromagnetic within each honeycomb strip.  This state is
connected to the HK-FM phase, and in the case of the \hhc{} lattice,
it is also connected to the $J>-K/2$ segment of the HK-SZ phase, where
ObD studies have shown that the HK-SZ phase orders in the $\hat{b}$
direction, much like the FM$_{a}$ phase.  The $C^a_2$ rotation
symmetry is preserved in this phase.

Since the $\hat{b}$ component of the phase in both lattices have a
vanishing net moment, only the $\hat{a}$ component contributes to the
total moment.  This total moment becomes saturated (spins point
entirely in the $\hat{a}$ direction) when approaching the HK-FM phase
and decreases smoothly as we approach the \ospa{-} boundary.  In the
\hhc{} case, the total moment further decreases and vanishes smoothly
as we approach the HK-SZ phase.  The magnitude of the moment along
$\hat{a}$ is depicted by the color contours in Fig. \ref{fig:pds}
where the largest projection is colored darkest.

\paragraph*{SS$_{x/y}$:}Wedged within the HK-SS, AF$_c$, and \ospb{+}
phases are two skew-stripy phases, the first of which to be discussed
is the non-coplanar SS$_{x/y}$ phase.  This
phase has the largest projection along the $x$($y$) direction and this
component orders in a skew-stripy fashion (these two orientations are
degenerate).  In the \hhc{} lattice, the other two Cartesian
components of the spins are small but finite and ensure that the spins
along each zigzag chain are collinear.  In the \har{} lattice, the
$y$($x$) component also orders in a skew-stripy fashion.  This phase
does not have a net moment and breaks all $C_2$ symmetries.

\paragraph*{SS$_b$:}This other skew-stripy phase is coplanar and lies
farther away from the FM Kitaev point relative to the SS$_{x/y}$
phase.  This phase borders the AF$_c$ phase and can be identified via
the LTA in the \hhc{} lattice.  In both lattices, the projection of
the spins along the $\hat{b}$ direction orders in a skew-stripy
pattern while the $\hat{c}$ projection vanishes.  In the \hhc{}
lattice, the $\hat{a}$ projection orders antiferromagnetically, while
in the \har{} lattice, spins order ferromagnetically along
non-bridging-$z$-bonds while spins along zigzag chains order
antiferromagnetically.  This state has a vanishing net moment and
preserves the $C^b_2$ rotation symmetry.

\paragraph*{SZ$_{x\negmedspace/\negmedspace y}$:}Analogous to the
SS$_{x/y}$ phase, there exists a non-coplanar, skew-zigzag phase with
a vanishing net moment near the AF Kitaev point bordering the \ospa{-}
phase that breaks all $C_2$ symmetries.  This phase has the largest
projection along the $x$($y$) direction and this component orders in a
skew-zigzag pattern (these two configurations are degenerate).  In
the \hhc{} lattice, the other two Cartesian spin components ensure
that the spins are collinear along zigzag chains.  In the \har{}
lattice, the $y$($x$) component is also ordered in the skew-zigzag
pattern.  This phase is closely related to the
SZ$_{x\negmedspace/\negmedspace y}$ phase: these two phases map onto
each other under the $(J,K,\Gamma)\rightarrow(-J,-K,-\Gamma)$
classical transformation.

\paragraph*{SZ$_b$:}In the \har{} model, an additional coplanar,
skew-zigzag phase with vanishing net moment exists.  This phase
borders the FM$_a$, SZ$_{x\negmedspace/\negmedspace y}$, \ospa{-}, and
HK-SZ phases.  It forms ferromagnetic skew-zigzag chains with
antiferromagnetic non-bridging-$z$ bonds and preserves the $C^a_2$
rotation symmetry.  The component along $\hat{c}$ vanishes while the
projection along $\hat{b}$ is the greatest.  This phase is closely
related to the SS$_b$ phase: these two phases map onto each other
under the $(J,K,\Gamma)\rightarrow(-J,-K,-\Gamma)$ classical
transformation.

\subsection{\label{subsec:mgamma}Phase diagrams for $\Gamma \le 0$ and
  connection to experimental results}

\begin{figure}[htbp!]
  \centering
  \setlength\fboxsep{0pt}
  \setlength\fboxrule{0pt}
  \subfloat[][\harzero{}: \ospabar{+}]{
    \label{fig:hhc_spa_p}
    \fbox{\begin{overpic}[scale=1,clip=true,trim=0 0 0 0]{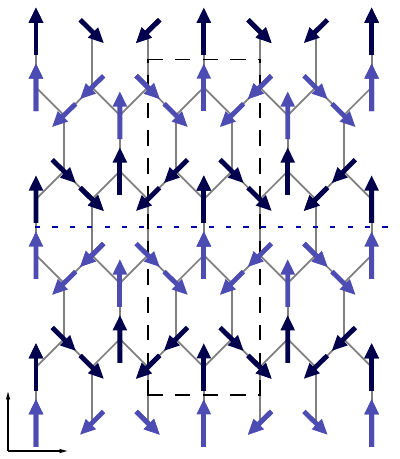}
        \put(0,21){\scriptsize{$c$}}
        \put(21,1){\scriptsize{$a$}}
        \put(-3,64){\scriptsize{$\overline{C}^a_2$}}
      \end{overpic}}
  }
  \subfloat[][\har{}: \ospabar{+}]{
    \label{fig:har_spa_p}
    \fbox{\begin{overpic}[scale=1,clip=true,trim=0 0 0 0]{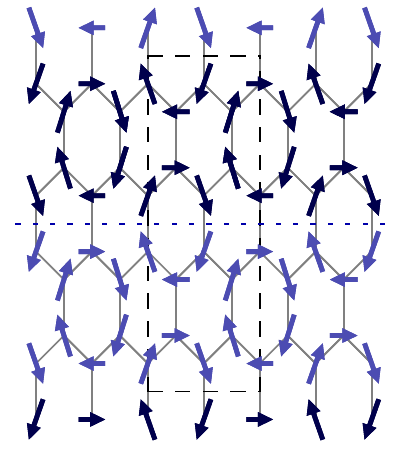}
      \end{overpic}}
  }

  \subfloat[][\harzero{}: \ospabar{-}]{
    \label{fig:hhc_spa_m}
    \fbox{\begin{overpic}[scale=1,clip=true,trim=0 0 0 0]{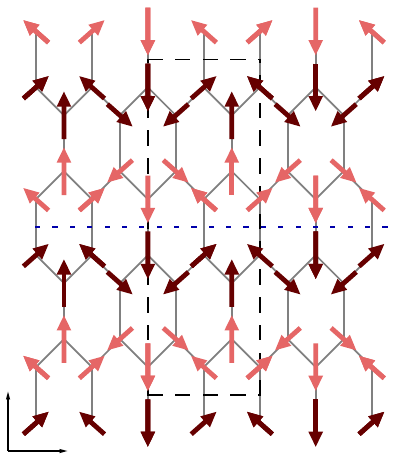}
        \put(0,21){\scriptsize{$c$}}
        \put(21,1){\scriptsize{$a$}}
        \put(-3,64){\scriptsize{$\overline{C}^a_2$}}
      \end{overpic}}
  }
  \subfloat[][\har{}: \ospabar{-}]{
    \label{fig:har_spa_m}
    \fbox{\begin{overpic}[scale=1,clip=true,trim=0 0 0 0]{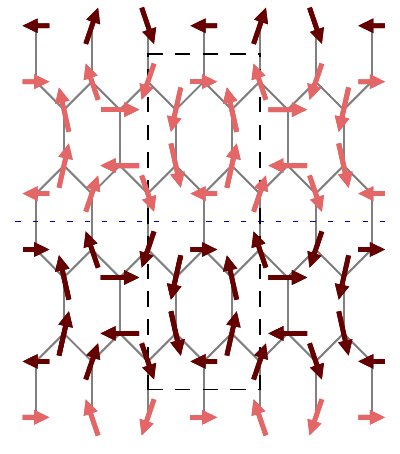}
      \end{overpic}}
  }
  \caption{\label{fig:realspace_a_mgamma}(Color online) Real-space
    spin configurations of the \ospabar{} spiral states obtained from
    simulated annealing with spins projected on to the $ac$-plane.
    The chosen parameter points yield $\mathbf{Q}=(0.33,0,0)$;
    however, we note that the $\mathbf{Q}$-vector in these phases are
    generally incommensurate and are not the same as the peak
    positions of the structure factor (see main text and
    Figs. \ref{fig:sf_a_mgamma}, \ref{fig:sf_b_mgamma}).  The black
    dashed box enclose the conventional unit cell.  Identical colors
    indicate that the sublattices share spiral-planes.  Shades of blue
    indicate that the spiral-planes are aligned with the
    honeycomb-planes while shades of red indicate that the
    spiral-planes are not aligned with the honeycomb-planes.  The
    handedness of adjacent sites can be readily verified as being
    opposite: the spirals counter-propagate.  Examples of the
    preserved rotation followed by time-reversal operation
    ($\overline{C}^a_2=\Theta \cdot C^a_2$ where $\Theta$ is the
    time-reversal operation) are indicated by the dotted blue lines.}
\end{figure}

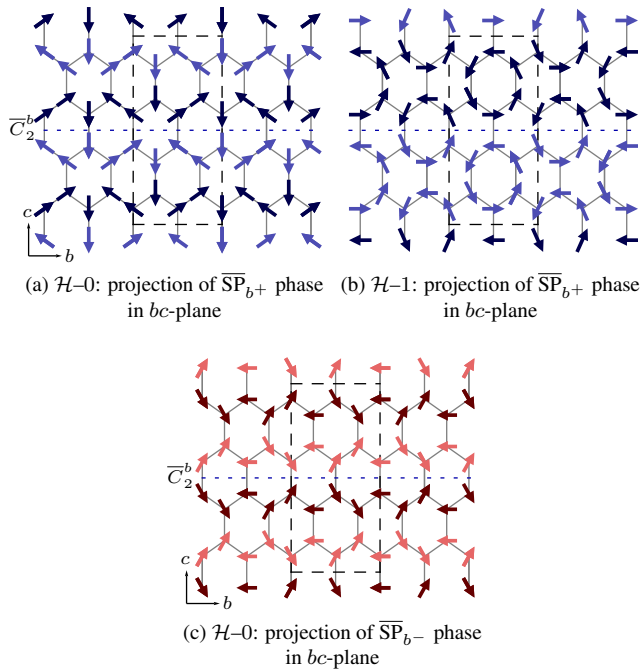
\begin{figure}[htbp!]
  \centering
  \setlength\fboxsep{0pt}
  \setlength\fboxrule{0pt}
  \subfloat[][\harzero{}: projection of \ospbbar{+} phase in $bc$-plane]{
    \label{fig:hhc_spb_p}
    \fbox{\begin{overpic}[scale=1,clip=true,trim=0 0 0 0]{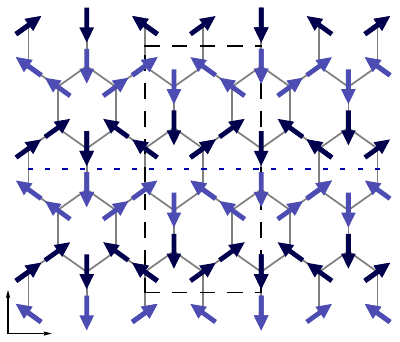}
        \put(0,17){\scriptsize{$c$}}
        \put(16,0){\scriptsize{$b$}}
        \put(-5,48){\scriptsize{$\overline{C}^b_2$}}
      \end{overpic}}
  }
  \subfloat[][\har{}: projection of \ospbbar{+} phase in $bc$-plane]{
    \label{fig:har_spb_p}
    \fbox{\begin{overpic}[scale=1,clip=true,trim=0 0 0 0]{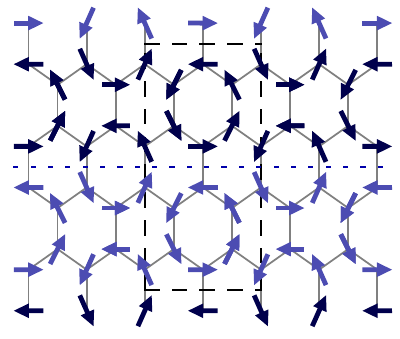}
      \end{overpic}}
  }

  \subfloat[][\harzero{}: projection of \ospbbar{-} phase in $bc$-plane]{
    \label{fig:hhc_spb_m}
    \fbox{\begin{overpic}[scale=1,clip=true,trim=0 0 0 0]{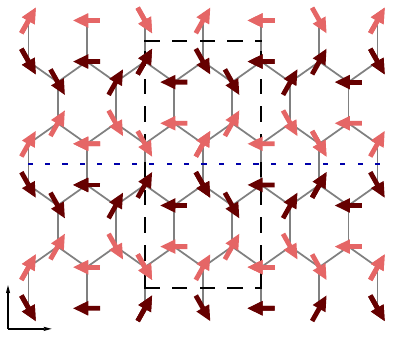}
        \put(0,17){\scriptsize{$c$}}
        \put(16,0){\scriptsize{$b$}}
        \put(-5,48){\scriptsize{$\overline{C}^b_2$}}
      \end{overpic}}
  }
  \caption{\label{fig:realspace_b_mgamma}(Color online) Real-space
    spin configurations of the \ospbbar{} spiral states obtained from
    simulated annealing with spins projected on to the $bc$-plane.
    The chosen parameter points yield $\mathbf{Q}=(0,0.33,0)$;
    however, we note that the $\mathbf{Q}$-vector in these phases are
    generally incommensurate.  See caption in
    Fig. \ref{fig:realspace_a_mgamma} for further details.
    $\overline{C}^b_2=\Theta\cdot C^b_2$ where $\Theta$ is the
    time-reversal operation.  In the case of the \ospbbar{-} phase,
    only the spiral planes and not the spins preserve the
    $\overline{C}^b_2$ operation.}
\end{figure}

As mentioned previously, the classical transformation of applying
time-reversal on the odd sublattices maps
$(J,K,\Gamma)\rightarrow(-J,-K,-\Gamma)$.  Using this transformation,
the classical phase diagrams for $\Gamma\ge 0$ in Fig. \ref{fig:pds}
can be inverted to yield the phase diagrams for $\Gamma\le 0$ in
Fig. \ref{fig:pds_neg}, where the radial coordinate now maps to
$r=\theta\in[\pi,\frac{\pi}{2}]$.  Under this transformation,
ferromagnetic components become antiferromagnetic, while stripy
components become zig-zag (these transformed phases are labeled in
Fig. \ref{fig:pds_neg}).  For the spiral and multiple-$\mathbf{Q}$
states, we use an overhead bar (\text{i.e.} \ospabar{-} versus
\ospa{-}) to emphasize the close relationship between the $\Gamma\le
0$ and the $\Gamma\ge 0$ phases.  For the spiral phases, we note that
the $\mathbf{Q}$-vector range remains invariant after this
transformation, while the $C_2$ rotation that the spiral state was
invariant under must now be followed by a time-reversal operation
$\Theta$.

We draw special attention to the \ospabar{-} phases found in the
$\Gamma\le 0$ region because of their close relationship to the
experimental magnetic orders (the real-space configuration of these
spiral phases can be seen in Fig. \ref{fig:realspace_a_mgamma} and can
be directly compared with those found in
Refs. \onlinecite{biffin2014non, biffin2014unconventional}).

The experimental magnetic ordering of the \hhc{}
\liiro{}\cite{biffin2014unconventional} possesses the same symmetries
as that of \ospabar{-}: following the analysis used in
Ref. \onlinecite{biffin2014unconventional}, we discern that the
\ospabar{-} phase on the \hhc{} lattice is described by the magnetic
basis vector combination $(iA_a,iC_b,F_c)$\footnote{The definintion of
  the $x,y,z$ axes in Refs. \onlinecite{biffin2014non,
    biffin2014unconventional} is along the orthorhombic directions.
  In this work, we have labeled these as the $\hat{a},\hat{b},\hat{c}$
  directions.  Hence, we have replaced $x,y,z$ with $a,b,c$ in the
  basis vector notation to avoid confusion.}, which is identical to
that of the experimental magnetic
ordering\cite{biffin2014unconventional}, indicating that these two
phases are indeed the same.  In addition, the experimental ordering
wavevector---$\mathbf{Q}_{\text{exp}}=(0.57,0,0)$---lies within the
range found in our model (when expressed using the same definition as
$\mathbf{Q}_{\text{exp}}$ from
Ref. \onlinecite{biffin2014unconventional}, the wavevectors found
within our present model is $(h'00)$ where $0.53\lesssim h' \lesssim
0.80$).

For the \har{} \liiro{}\cite{biffin2014non}, $\mathbf{Q}_{\text{exp}}$
also lies within the range found in our model.  Using the notation
defined in Ref. \onlinecite{biffin2014non}, the magnetic basis vector
combination that describes the \ospabar{-} phase is
$[i(A,-A)_a,-i(C,-C)_b, (F,F)_c]$, which only differs from the
experimental combination of $[i(A,-A)_a, (-1)^mi(F,-F)_b,(F,F)_c]$ in
the $S_b$-component.

For completeness, we also show the real-space configurations of the
\ospabar{+} and \ospbbar{} phases in Figs. \ref{fig:realspace_a_mgamma}
and \ref{fig:realspace_b_mgamma} respectively, as a comparison with
Figs. \ref{fig:realspace_a} and \ref{fig:realspace_b}.  Magnetic static
structure factors of these spiral phases can be found in Appendix
\ref{app:magneticstructurefactor}.

\section{\label{sec:discussion}Discussion and Outlook}
\begin{table*}[htbp!]
  \caption{\label{tbl:orders}Summary of magnetic phases for
    $\Gamma \ge 0$.  For detailed description of each phase,
    see Sec. \ref{sec:orders}.  Magnetic phases for $\Gamma \le 0$ are
    obtained by acting the time-reversal operation on odd sublattices
    on the phases summarized here; see Sec. \ref{subsec:mgamma} for details.
    Under this classical transformation,
    the $\mathbf{Q}$-vector range of the spiral phases remain invariant,
    while the preserved $C_2$ symmetry must now be followed by the time-reversal
    operation $\Theta$, \textit{e.g.} the transformed spiral states \ospabar{+} are
    invariant under $\overline{C}^a_2\equiv\Theta \cdot C^a_2$.}
\begin{ruledtabular}
\begin{tabular}{c|cccccccc}
  & \multirow{2}{*}{$S_{\text{tot}}$\footnotemark[1]} & \multirow{2}{*}{$\mathcal{H}\text{--}0$\footnotemark[2]} & \multirow{2}{*}{\har{}\footnotemark[2]} & Collinear/ & Preserved & Structure & $\mathbf{Q}$-vector &\multirow{2}{*}{Notes} \\
  & & & & coplanar & symmetry & factor peak ($\mathbf{q}$)\footnotemark[3] & range\footnotemark[4] &\\
  \hline
  \ospa{+} &  & NM & NM & & $C^a_2$ & $(\mathbf{1}_a+\mathbf{Q})$& \parbox[t]{1.3in}{$\sim(0.33,0,0)$}&\parbox[t]{1.5in}{spiral-planes aligned \\ with honeycomb-planes}\\
  \ospb{+} &  & NM & NM & & $C^b_2$ & $(\mathbf{1}_b+\mathbf{Q})$& \parbox[t]{1.3in}{$(0k0)$\\$\mathcal{H}\text{--}0$: $0.17\lesssim k\lesssim0.40$\\\har{}: $0.30\lesssim k\lesssim0.36$}&\parbox[t]{1.5in}{spiral-planes aligned \\ with honeycomb-planes}\\
  \ospa{-} &  & NM & NM & & $C^a_2$ & $(\mathbf{1}_a-\mathbf{Q})$& \parbox[t]{1.3in}{$(h00)$\\$\mathcal{H}\text{--}0$: $0.20\lesssim h\lesssim0.47$\\\har{}: $0.15\lesssim h\lesssim0.43$}&\parbox[t]{1.5in}{spiral-planes not aligned \\ with honeycomb-planes}\\
  \ospb{-} &  & NM & n/a & & $C^b_2$\footnotemark[5] & $(\mathbf{1}_b-\mathbf{Q})$& \parbox[t]{1.3in}{$\sim(0,0.33,0)$}&\parbox[t]{1.5in}{spiral-planes not aligned \\ with honeycomb-planes}\\
  \hline
  MQ$_1$ &  & NM & NM & & & & $(0.5,0.5,x)$ & multiple-$q$ state\\
  MQ$_2$ & $\hat{a}$ & NM & NM & & & & \parbox[t]{1.3in}{$(\pm0.33,\pm0.33,0)$,\\$(\pm0.33,0,0)$,$(0,0,0)$}& \parbox[t]{1.5in}{multiple-$\mathbf{Q}$ state \\ with finite moment}\\
  AF-SZ$_{\text{AF}}$ & & LTA & NM & copl. & $C^a_2$ & & & \parbox[t]{1.5in}{AF skew-zigzag chains}\\
  FM-SZ$_{\text{AF}}$ & $\hat{a}$ & n/a & NM & copl. & $C^a_2$ & & &\parbox[t]{1.5in}{AF skew-zigzag chains,\\FM non-bridging-$z$-bonds}\\
  \hline
  AF$_{c}$ &  & LTA & LTA & col. & $C^a_2$/$C^b_2$/$C^c_2$& & &\parbox[t]{1.5in}{AF order along $\hat{c}$}\\
  FM$_{a}$ & $\hat{a}$ & LTA & NM & copl. & $C^a_2$ & & &\parbox[t]{1.5in}{$S_i\cdot\hat{a}$: FM order}\\
  SS$_{b}$ &  & LTA & NM & copl. & $C^b_2$ & & &\parbox[t]{1.5in}{$S_i\cdot\hat{b}$: SS order}\\
  SZ$_{b}$ &  & n/a & NM & copl. & $C^a_2$ & & &\parbox[t]{1.5in}{$S_i\cdot\hat{b}$: SZ order}\\
  SS$_{x\negmedspace/\negmedspace y}$ &  & NM & NM & & & & &\parbox[t]{1.5in}{$S_i\cdot\hat{x}$ or $S_i\cdot\hat{y}$: SS order (degenerate)}\\
  SZ$_{x\negmedspace/\negmedspace y}$ &  & NM & NM & & & & &\parbox[t]{1.5in}{$S_i\cdot\hat{x}$ or $S_i\cdot\hat{y}$: SZ order (degenerate)}\\
\end{tabular}
\end{ruledtabular}
\footnotetext[1]{Indicates the direction of the total moment, if finite}
\footnotetext[2]{Method that the ground state was determined: Luttinger-Tisza approximation (LTA) or numerical minimization via single-$\mathbf{Q}$ ansatz and simulated annealing (NM); a blank entry is assigned if the phase does not exist for the given model}
\footnotetext[3]{$\mathbf{1}_a=(100)$ and $\mathbf{1}_b=(010)$.}
\footnotetext[4]{Using the alternative definition of the ordering wavevector in Refs. \onlinecite{biffin2014non, biffin2014unconventional} (hereby denoted as $\mathbf{Q}'$), the spiral phases considered in this work are characterized by $\mathbf{Q}'=(\mathbf{1}_x-\mathbf{Q})$, \textit{i.e.} for the \ospa{-} phase of the \hhc{} lattice,  $\mathbf{Q}'=(h'00)$ where $0.53\lesssim h'\lesssim0.80$.}
\footnotetext[5]{Only the spiral-planes and not the spins preserve the rotation symmetry}
\end{table*}

In the model presented, we have implicitly assumed an ideal crystal
structure.  In principle, distortions of the oxygen octahedra as well
as Ir positions can bring about additional anisotropic exchange
interactions between neighboring pseudospins.\cite{katukuri2014kitaev,
  yamaji2014honeycomb} In addition, large distortions can admix the
\jhalf{} and the neglected orbitals, which can generate anisotropic
$g$-factors or even destroy the \jhalf{} description
entirely.\cite{bhattacharjee2012spin} Nevertheless, these distortion
effects are likely to be smaller in the 3D compounds compared to the
2D honeycomb iridates, since the octahedral environments in the former
are more ideal than those of the latter.\cite{choi2012spin,
  takayama2014hyper, modic2014realization, manni2014effect}.  As such,
these distortion effects may serve only to affect quantitatively the
phases found within our treatment while the qualitative features of
our phase diagrams can be expected to remain robust.

We have also assumed that the 3D materials lie deep in the Mott
insulating regime ($U\rightarrow \infty$) such that other hopping
pathways, namely ones involving multiple oxygen and lithium ions, can
be neglected, thus resulting in a model with only NN exchanges.  The
role of further neighbors have been considered in the 2D honeycomb
iridates: further neighbor hopping in \textit{ab initio}
treatments\cite{mazin2012na, kim2012topological, foyevtsova2013ab,
  kim2013strain}, as well as further neighbor exchanges in the
localized picture\cite{kimchi2011kitaev, reuther2014spiral} have been
investigated.  In addition, experimental work has suggested the
possibility of significant further neighbor interactions in the case
of the 2D \liiro{}.\cite{manni2014effectof} Meanwhile, there have also
been studies on effects of further neighbors in the case of the \hhc{}
iridate in both the itinerant picture\cite{lee2014topological} and in
the Mott limit\cite{lee2014emergent}.  The effects of further
neighbors have revealed interesting consequences in these cases, which
suggests that additional studies on this subject and its role in a
minimal model for the \hhc{} and \har{} lattices may be warranted.

On the other hand, if the ideal NN localized picture is adequate in
capturing the magnetic order of the whole family of honeycomb
iridates, then we may expect the 2D honeycomb, \hhc{}, and \har{}
iridates to be describable by NN $J\text{--}K\text{--}\Gamma$ models.
The 2D-\liiro{} and the two 3D-\liiro{} are then expected to have
comparable exchange parameters, since their chemical composition and
local crystal structures are similar, while the 2D-Na$_2$IrO$_3$ is
more distorted and may lie in a different region in parameter space.
Since the 2D-Na${}_{2}$IrO${}_{3}$ orders in a zigzag
fashion\cite{liu2011long, ye2012direct, choi2012spin} and the
2D-\liiro{} is believed to order in a spiral
phase\cite{reuther2014spiral, coldea2014}, this line of reasoning
suggests that the 3D-\liiro{}'s may order in one of the spiral phases,
especially since the spiral regions are relatively larger and the
zigzag region smaller in the 3D phase diagrams.  Indeed, the two
3D-\liiro{} have been experimentally reported to order in a
non-collinear fashion, with counter-propagating
spirals\cite{biffin2014non,biffin2014unconventional}.  The ordering
pattern in the case of the \hhc{} \liiro{} case is precisely that of
the \hhc{} \ospabar{-} phase\cite{biffin2014unconventional}, while
that of the \har{} \liiro{} has striking similarities with the \har{}
\ospabar{-} phase\cite{biffin2014non}.  Moreover, the similar
thermodynamic properties of the two 3D-\liiro{} lattices may be due to
the combination of the near-identical phase diagrams and the fact that
both systems are near ideal (distortion-free) and largely driven by NN
physics.

In summary, motivated by the similar behaviors and local structure of
the \hhc{} and \har{} \liiro{}, we studied the NN pseudospin
$J\text{-}K\text{-}\Gamma$ model, which contains the Heisenberg, the
Kitaev, and the off-diagonal $\Gamma$ exchange interactions. We argue
that with strong on-site interaction and atomic spin-orbit coupling,
these models are minimal effective models for the \jhalf{} pseudospins
in these 3D honeycomb-based materials.  The resulting classical phase
diagrams that emerge in the two cases are very similar.  On the other
hand, when compared with the phase diagram of the 2D honeycomb
lattice, the zigzag region diminished while spiral phases with
counter-propagating spirals expanded in the 3D phase diagrams.  We
characterized all the magnetic orders in detail and provided static
structure factor results of the spiral phases.  We believe that our
phase diagrams and detailed analysis will be relevant in identifying
the phases of the 3D-\liiro{} compounds, and our model can serve as
the minimal starting point in identifying the common thread which runs
through this family of iridium-based materials.

\acknowledgments We thank R. Schaffer and J. Rau for discussions.  We
also thank R. Coldea for his useful comments on our work and for
making some of the results in Refs. \onlinecite{biffin2014non} and
\onlinecite{biffin2014unconventional} available prior to publication.
Computations were performed on the GPC supercomputer at the SciNet HPC
Consortium. SciNet is funded by: the Canada Foundation for Innovation
under the auspices of Compute Canada; the Government of Ontario;
Ontario Research Fund - Research Excellence; and the University of
Toronto.  This research was supported by the NSERC, CIFAR, and Centre
for Quantum Materials at the University of Toronto.

Upon initial submission of this manuscript to the pre-print arXiv,
independent experimental results on the magnetic structure of \har{}
\liiro{} were released\cite{biffin2014non} which reported the
existence of a spiral order in the compound.  In addition, independent
experimental report of the existence of a spiral order in \hhc{}
\liiro{} was released\cite{biffin2014unconventional} after the
submission of this work.  Although this work is an exploratory study
of our pseudospin model, we have since made modifications to reference
these experimental results throughout the current manuscript wherever
relevant.

\appendix
\section{\label{app:model}Details on the strong-coupling expansion}
Octahedral crystal fields arising from the oxygen ions split the \ir{}
$5d$ orbitals into the lower energy $t_{2g}$ and higher energy $e_{g}$
multiplets.  Large atomic spin-orbit coupling further split the
$t_{2g}$ orbitals into the lowest energy \jthreehalf{} quadruplet and
higher energy \jhalf{} doublets.  The five electrons at each \ir{}
site fully-fill the \jthreehalf{} orbitals and half-fill the \jhalf{}
orbitals, leaving the $e_{g}$ orbitals unoccupied.  In the limit of
large atomic spin-orbit coupling and strong on-site electron
correlations, the \jthreehalf{} and $e_{g}$ orbitals can be projected
out and the relevant low-energy degrees of freedom are the localized
\jhalf{} pseudospins.

In the presence of hopping between neighboring Ir orbitals, the
perturbed ground states can be captured by an effective Hamiltonian
derived by a strong coupling expansion.  We assume the unperturbed,
atomic Hamiltonian to be comprised of atomic SOC and on-site
interactions.  The atomic SOC takes the form $H_{\text{SOC}}=-\sum_{a}
\lambda l_{a} \cdot s_{a}$, where $s_{a}$ and $l_{i}$ are the spin and
effective $t_{2g}$ orbital angular momentum of electron $a$.  The
on-site interaction consists of Hund's coupling and
intra/inter-orbital repulsion, which we write in the Kanamori
form
\begin{equation}
  H_{\text{on-site}}=\sum_{i}\left[ \frac{U-3J_{H}}{2} \left(N_i - 5\right)^2 - 2 J_{H}S^2_{i} - \frac{J_{H}}{2} L^2_i \right]
\end{equation}
where $N_i$, $S_i$, and $L_i$ are the total number, spin, and
effective orbital angular momentum at site $i$; $U$ and $J_H$ are the
Coulomb repulsion and Hund's coupling respectively.  The perturbation
comes in the form of NN hopping amplitudes between $t_{2g}$ orbitals:
we introduce both direct orbital overlaps and effective hopping
between neighboring orbitals mediated by the edge-shared oxygens.  The
hopping amplitudes can be written as
\begin{align}
  H_{\text{hop}}=\sum_{\langle i,j\rangle \in \alpha \beta (\gamma)}&\left[
  t_1\left(d^{\dagger}_{i\alpha}d_{j\alpha}+d^{\dagger}_{i\beta}d_{j\beta}\right) \right.\nonumber\\
  &\left.+t_2\left(d^{\dagger}_{i\alpha}d_{j\beta}+d^{\dagger}_{i\beta}d_{j\alpha}\right)
  +t_3d^{\dagger}_{i\gamma}d_{j\gamma} \right],
\end{align}
where $\gamma\in(x,y,z)$ is the label for bond $\langle i,j \rangle$,
and $\alpha, \beta$ are the other two components.  The creation
operator is given by
$d^{\dagger}_{i\alpha}=\left(d^{\dagger}_{i\alpha\uparrow},d^{\dagger}_{i\alpha\downarrow}\right)$
and creates electrons on site $i$ in orbital $\alpha$, where the
$t_{2g}$ orbitals are labeled according to $x\rightarrow yz$,
$y\rightarrow xz$, and $z\rightarrow xy$.  The hopping amplitudes are given by
\begin{equation}
  t_1=\frac{t_{dd\pi}+t_{dd\delta}}{2}, t_2=\frac{t^2_{pd\pi}}{\Delta}+\frac{t_{dd\pi}-t_{dd\delta}}{2}, t_3=\frac{3t_{dd\sigma}+t_{dd\delta}}{4}
\end{equation}
where $t_{dd\sigma}$, $t_{dd\pi}$, $t_{dd\delta}$, and $t_{pd\pi}$ are
the Slater-Koster amplitudes parametrizing direct Ir orbital overlaps
and O-Ir orbital overlaps respectively, while $\Delta$ is the energy
gap between oxygen and iridium.

In the large SOC limit, we perform the strong coupling expansion with
the limit $U > \lambda \gg J_H \gg t$ then project into the \jhalf{}
manifold.  The resulting \jhalf{} pseudospin Hamiltonian then takes
the form of Eq. \ref{eq:ham} with exchanges related to hopping
amplitudes by Eq. \ref{eq:exchanges}.  For more detail on the
derivation of the expansion, we refer the reader to the appendix in
Ref. \onlinecite{rau2014generic}.

From Eq. \ref{eq:exchanges}, we note that the exchanges depend
sensitively on the relative signs and magnitudes of the hoping
amplitudes $t_{1-3}$.  Hence, a detailed analysis of the underlying
band structure would greatly aid the realistic estimations of these
exchanges for the 3D \liiro{} systems.

\section{\label{app:sign}Sign structure of $\Gamma$}

\begin{figure}[htbp!]
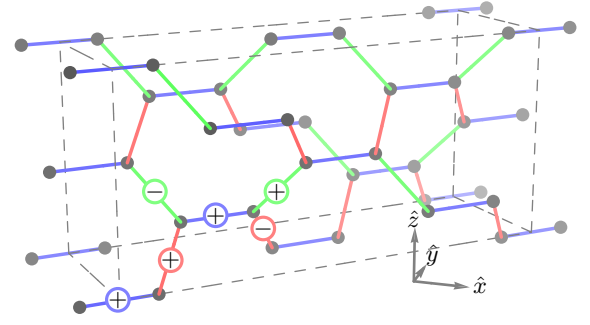
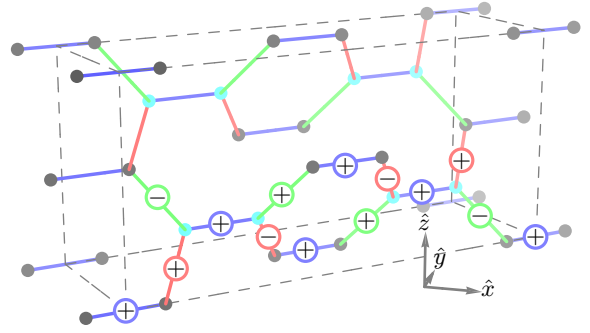

  \centering
  \setlength\fboxsep{0pt}
  \setlength\fboxrule{0pt}
  \subfloat[][Hyperhoneycomb lattice]{
    \fbox{\begin{overpic}[scale=1,clip=true,trim=15 0 -15 0]{hhc_lattice}
        \put(45.7,21.5){\bdot{}}
        \put(47,24){$+$}
        \put(82.7,53.5){\bdot{}}
        \put(84,56){$+$}
        \put(65.7,36.5){\rdot{}}
        \put(67,39){$+$}
        \put(100.7,48.5){\rdot{}}
        \put(102,51){$-$}
        \put(59.7,62.5){\gdot{}}
        \put(61,65){$-$}
        \put(105.7,62.5){\gdot{}}
        \put(107,65){$+$}
        \put(185,29){$\hat{x}$}
        \put(167,41){$\hat{y}$}
        \put(160,54){$\hat{z}$}
      \end{overpic}}
  }

  \subfloat[][\har{} lattice]{
    \fbox{\begin{overpic}[scale=1,clip=true,trim=15 0 -15 0]{har_lattice}
        \put(48.7,18.5){\bdot{}}
        \put(50,21){$+$}
        \put(84.7,51.5){\bdot{}}
        \put(86,54){$+$}
        \put(116.7,41.5){\bdot{}}
        \put(118,44){$+$}
        \put(131.7,73.5){\bdot{}}
        \put(133,76){$+$}
        \put(160.7,63.5){\bdot{}}
        \put(162,66){$+$}
        \put(203.7,46.5){\bdot{}}
        \put(205,49){$+$}
        \put(67.7,34.5){\rdot{}}
        \put(69,37){$+$}
        \put(102.7,46.5){\rdot{}}
        \put(104,49){$-$}
        \put(147.7,68.5){\rdot{}}
        \put(149,71){$-$}
        \put(175.7,75.5){\rdot{}}
        \put(177,78){$+$}
        \put(60.7,61.5){\gdot{}}
        \put(62,64){$-$}
        \put(107.7,62.5){\gdot{}}
        \put(109,65){$+$}
        \put(139.7,52.5){\gdot{}}
        \put(141,55){$+$}
        \put(182.7,54.5){\gdot{}}
        \put(184,57){$-$}
        \put(188,29){$\hat{x}$}
        \put(170,41){$\hat{y}$}
        \put(164,54){$\hat{z}$}
      \end{overpic}}
  }
  \caption{\label{fig:sign_lattice}(Color online) Relative signs of
    $\Gamma$ chosen for the \hhc{} and \har{} lattices.  The
    plus/minus sign on each bond corresponds to the sign in front of
    $\Gamma$ that appears in the Hamiltonian in Eq. \ref{eq:ham}.
    With this choice of relative signs in $\Gamma$, we are still free
    to choose either a positive or a negative value for $\Gamma$.  The
    $x$, $y$, and $z$ bonds are colored red, green, and blue,
    respectively.  The bridging-sites are colored cyan.  The gray
    dashed boxes are the conventional unit cells.}
\end{figure}

Although the relative signs of $\Gamma$ among
symmetry-\textit{equivalent} bonds can be determined by applying the
appropriate symmetry transformation, the relative signs among
symmetry-\textit{inequivalent} bonds are not constrained by crystal
symmetries.  However, these symmetry-inequivalent bonds have identical
local environments (\textit{e.g.} nearest-neighbor oxygen and cation
hopping pathways), and hence, the sign structure of $\Gamma$ among
symmetry-inequivalent bonds are expected to follow transformations
that preserve these local environments.  For example, to relate the
signs of $\Gamma$ on the $x$- and $z$-bonds emanating from a site, a
$C_2$ rotation through the site and along the connecting $y$-bond
would preserve the local oxygen and cation environments while mapping
the $x$-bond to the $z$-bond. This microscopic justification of the
relative signs of $\Gamma$ was used in assigning the sign structures
in Fig. \ref{fig:sign_lattice}.  We emphasize that this convention
only fixes the \textit{relative} signs $\Gamma$ in the lattice:
$\Gamma$ can still be either positive or negative, leading to the
phase diagrams in Figs. \ref{fig:pds} and \ref{fig:pds_neg}
respectively.  We leave other possible choices of relative signs in
$\Gamma$ among symmetry-inequivalent bonds for future study.

\section{\label{app:magneticstructurefactor}Magnetic structure factors}
\begin{figure}[htbp!]
  \centering
  \setlength\fboxsep{0pt}
  \setlength\fboxrule{0pt}
  \subfloat[][\harzero{}: \ospa{+}]{
    \label{fig:hhc_spa_p_sf}
    \fbox{\begin{overpic}[scale=1,clip=true,trim=0 -10 0 0]{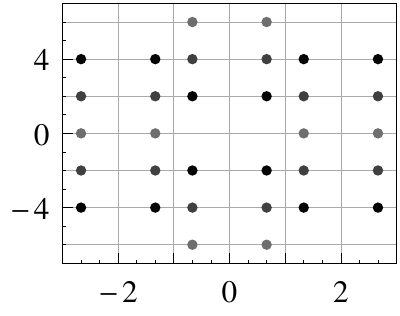}
        \put(0,59){l}
        \put(64,5){h}
      \end{overpic}}
  }
  \subfloat[][\har{}: \ospa{+}]{
    \label{fig:har_spa_p_sf}
    \fbox{\begin{overpic}[scale=1,clip=true,trim=0 -10 0 0]{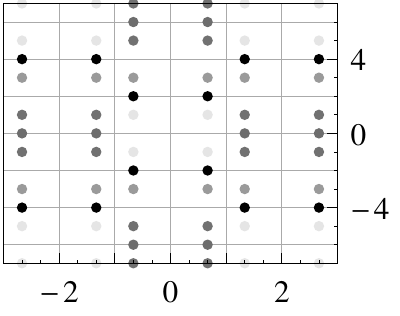}
        \put(47,5){h}
      \end{overpic}}
  }

  \subfloat[][\harzero{}: \ospa{-}]{
    \label{fig:hhc_spa_m_sf}
    \fbox{\begin{overpic}[scale=1,clip=true,trim=0 -10 0 0]{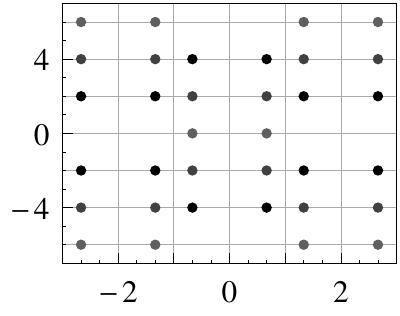}
        \put(0,59){l}
        \put(64,5){h}
      \end{overpic}}
  }
  \subfloat[][\har{}: \ospa{-}]{
    \label{fig:har_spa_m_sf}
    \fbox{\begin{overpic}[scale=1,clip=true,trim=0 -10 0 0]{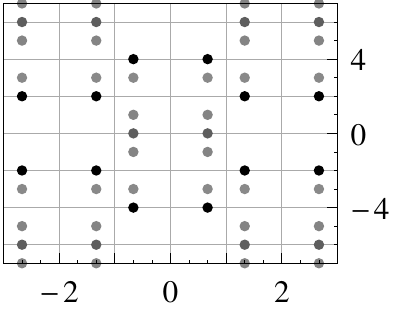}
        \put(47,5){h}
      \end{overpic}}
  }
  \caption{\label{fig:sf_a} Static structure factors along the (h0l)
    plane for the \ospa{} phases presented in
    Fig. \ref{fig:realspace_a}.  These states are characterized by
    $\mathbf{Q}=(0.33,0,0)$.  The darkness of the dots indicate the
    peaks' relative intensities.  The $+$ spiral phases have peaks at
    $\mathbf{q}=(\mathbf{1}_a+\mathbf{Q})=(1.33,0,0)$ while the $-$
    spiral phases have peaks at
    $\mathbf{q}=(\mathbf{1}_a-\mathbf{Q})=(0.66,0,0)$, where
    $\mathbf{1}_a=(100)$.}
\end{figure}

\begin{figure}[htbp!]
  \centering
  \setlength\fboxsep{0pt}
  \setlength\fboxrule{0pt}
  \subfloat[][\harzero{}: \ospb{+}]{
    \label{fig:hhc_spb_p_sf}
    \fbox{\begin{overpic}[scale=1,clip=true,trim=0 -10 0 0]{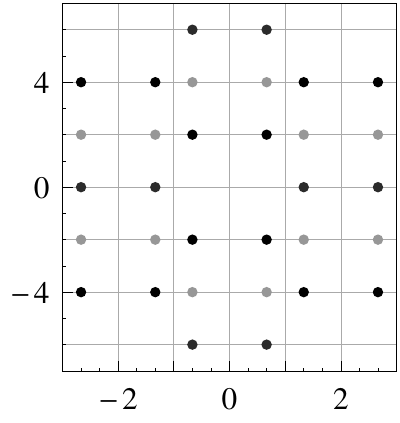}
        \put(0,75){l}
        \put(64,5){k}
      \end{overpic}}
  }
  \subfloat[][\har{}: \ospb{+}]{
    \label{fig:har_spb_p_sf}
    \fbox{\begin{overpic}[scale=1,clip=true,trim=0 -10 0 0]{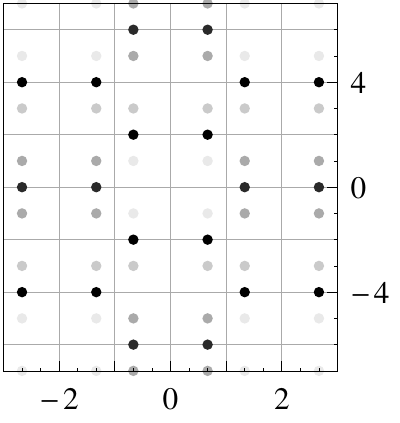}
        \put(47,5){k}
      \end{overpic}}
  }

  \subfloat[][\harzero{}: \ospb{-}]{
    \label{fig:hhc_spb_m_sf}
    \fbox{\begin{overpic}[scale=1,clip=true,trim=0 -10 0 0]{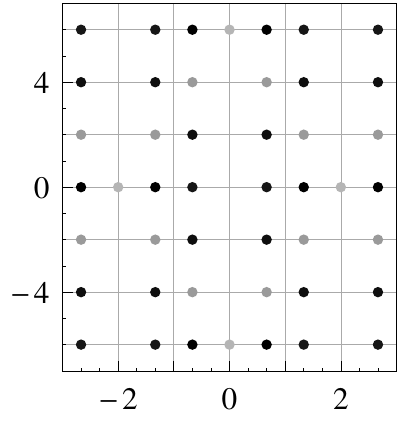}
        \put(0,75){l}
        \put(64,5){k}
      \end{overpic}}
  }
  \caption{\label{fig:sf_b} Static structure factors along the (0kl)
    plane for the \ospb{} phases presented in
    Fig. \ref{fig:realspace_b}.  These states are characterized by
    $\mathbf{Q}=(0,0.33,0)$.  The darkness of the dots indicate the
    peaks' relative intensities.  The $+$ spiral phases have peaks at
    $\mathbf{q}=(\mathbf{1}_b+\mathbf{Q})=(0,1.33,0)$ while the $-$ spiral phase
    has a peak at $\mathbf{q}=(\mathbf{1}_b-\mathbf{Q})=(0,0.66,0)$, where
    $\mathbf{1}_b=(010)$.}
\end{figure}

\begin{figure}[htbp!]
  \centering
  \setlength\fboxsep{0pt}
  \setlength\fboxrule{0pt}
  \subfloat[][\harzero{}: \ospabar{+}]{
    \label{fig:hhc_spa_p_sf}
    \fbox{\begin{overpic}[scale=1,clip=true,trim=0 -10 0 0]{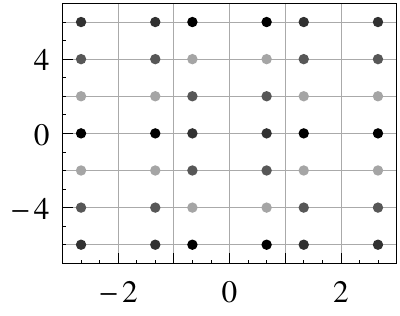}
        \put(0,59){l}
        \put(64,5){h}
      \end{overpic}}
  }
  \subfloat[][\har{}: \ospabar{+}]{
    \label{fig:har_spa_p_sf}
    \fbox{\begin{overpic}[scale=1,clip=true,trim=0 -10 0 0]{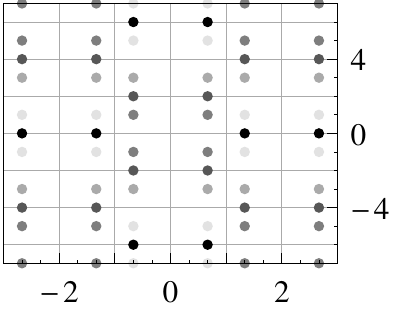}
        \put(47,5){h}
      \end{overpic}}
  }

  \subfloat[][\harzero{}: \ospabar{-}]{
    \label{fig:hhc_spa_m_sf}
    \fbox{\begin{overpic}[scale=1,clip=true,trim=0 -10 0 0]{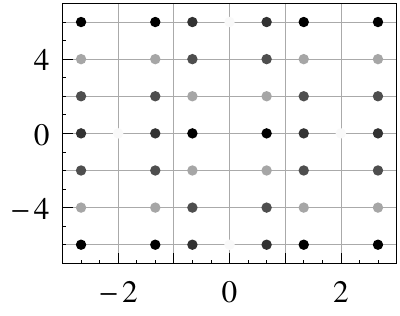}
        \put(0,59){l}
        \put(64,5){h}
      \end{overpic}}
  }
  \subfloat[][\har{}: \ospabar{-}]{
    \label{fig:har_spa_m_sf}
    \fbox{\begin{overpic}[scale=1,clip=true,trim=0 -10 0 0]{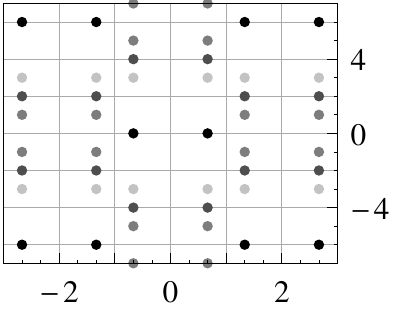}
        \put(47,5){h}
      \end{overpic}}
  }
  \caption{\label{fig:sf_a_mgamma} Static structure factors along the
    (h0l) plane for the $\Gamma\le0$ \ospabar{} phases presented in
    Fig. \ref{fig:realspace_a_mgamma}.  These states are characterized
    by $\mathbf{Q}=(0.33,0,0)$.  The darkness of the dots indicate the
    peaks' relative intensities.}
\end{figure}

\begin{figure}[htbp!]
  \centering
  \setlength\fboxsep{0pt}
  \setlength\fboxrule{0pt}
  \subfloat[][\harzero{}: \ospbbar{+}]{
    \label{fig:hhc_spb_p_sf}
    \fbox{\begin{overpic}[scale=1,clip=true,trim=0 -10 0 0]{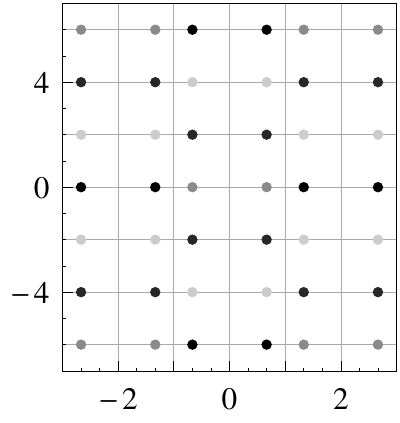}
        \put(0,75){l}
        \put(64,5){k}
      \end{overpic}}
  }
  \subfloat[][\har{}: \ospbbar{+}]{
    \label{fig:har_spb_p_sf}
    \fbox{\begin{overpic}[scale=1,clip=true,trim=0 -10 0 0]{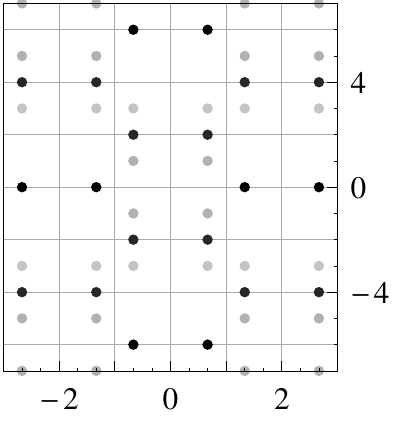}
        \put(47,5){k}
      \end{overpic}}
  }

  \subfloat[][\harzero{}: \ospbbar{-}]{
    \label{fig:hhc_spb_m_sf}
    \fbox{\begin{overpic}[scale=1,clip=true,trim=0 -10 0 0]{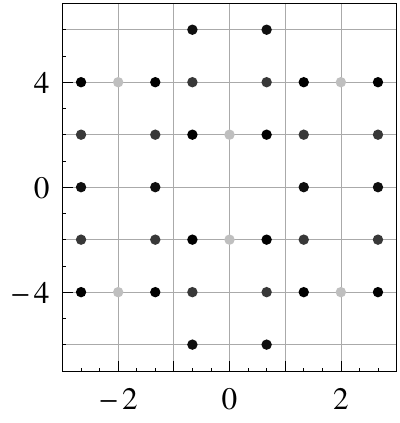}
        \put(0,75){l}
        \put(64,5){k}
      \end{overpic}}
  }
  \caption{\label{fig:sf_b_mgamma} Static structure factors along the (0kl)
    plane for the $\Gamma\le0$ \ospbbar{} phases presented in
    Fig. \ref{fig:realspace_b_mgamma}.  These states are characterized by
    $\mathbf{Q}=(0,0.33,0)$.  The darkness of the dots indicate the
    peaks' relative intensities.}
\end{figure}

The magnetic static structure factors along certain momentum cuts of
the \ospa{}, \ospb{}, \ospabar{}, and \ospbbar{} phases of both the
\hhc{} and \har{} lattices are presented in Figs. \ref{fig:sf_a},
\ref{fig:sf_b}, \ref{fig:sf_a_mgamma}, and \ref{fig:sf_b_mgamma}
respectively for reference.

\end{document}